\documentclass{PoS}

\usepackage{wrapfig}
\usepackage{amsmath}
\usepackage{amsfonts}
\usepackage{slashed}

\usepackage{bbold}

\newcommand{\dd}{\mathrm{d}}
\DeclareMathAlphabet{\mathdsl}{U}{bbm}{m}{sl}
\usepackage{arydshln}
\usepackage{colortbl}

\DeclareMathAlphabet{\mathbbmsl}{U}{bbm}{m}{sl}
\makeatletter
\DeclareFontFamily{OMX}{MnSymbolE}{}
\DeclareSymbolFont{MnLargeSymbols}{OMX}{MnSymbolE}{m}{n}
\SetSymbolFont{MnLargeSymbols}{bold}{OMX}{MnSymbolE}{b}{n}
\DeclareFontShape{OMX}{MnSymbolE}{m}{n}{
    <-6>  MnSymbolE5
   <6-7>  MnSymbolE6
   <7-8>  MnSymbolE7
   <8-9>  MnSymbolE8
   <9-10> MnSymbolE9
  <10-12> MnSymbolE10
  <12->   MnSymbolE12
}{}
\DeclareFontShape{OMX}{MnSymbolE}{b}{n}{
    <-6>  MnSymbolE-Bold5
   <6-7>  MnSymbolE-Bold6
   <7-8>  MnSymbolE-Bold7
   <8-9>  MnSymbolE-Bold8
   <9-10> MnSymbolE-Bold9
  <10-12> MnSymbolE-Bold10
  <12->   MnSymbolE-Bold12
}{}

\let\llangle\@undefined
\let\rrangle\@undefined
\DeclareMathDelimiter{\llangle}{\mathopen}%
                     {MnLargeSymbols}{'164}{MnLargeSymbols}{'164}
\DeclareMathDelimiter{\rrangle}{\mathclose}%
                     {MnLargeSymbols}{'171}{MnLargeSymbols}{'171}

\newcommand{\tikzmarkMath}[2]{%
    \tikz[%
        remember picture, 
        baseline = (#1.base),
        inner sep = 0pt,
        outer sep = 0pt,
    ] \node (#1) {$\m@th\displaystyle #2$};%
}
\makeatother

\usepackage{tikz}
\usetikzlibrary{shapes,arrows,calc}
\usepackage{cite}

\newcommand{\DFTwzw}{DFT${}_\mathrm{WZW}$}

\title{An invitation to Poisson-Lie T-duality in\\ Double Field Theory and its applications}
\ShortTitle{PL T-duality and Double Field Theory}
\author{Saskia Demulder\\
        Theoretische Natuurkunde, Vrije Universiteit Brussel \& The
        International Solvay Institutes,\\ B-1050 Brussels, Belgium,\\
         Department of Physics, Swansea University, Swansea, SA2 8PP, U.K. \\
        E-mail: \email{Saskia.Demulder@vub.be}}

\author{\speaker{Falk Hassler}\\
        Department of Physics, University of Oviedo, Avda. Federico Garc\'ia Lorca 18,
        33007 Oviedo, Spain.\\
        E-mail: \email{falk@fhassler.de}}

\author{Daniel C. Thompson\\
  Theoretische Natuurkunde, Vrije Universiteit Brussel \& The
        International Solvay Institutes,\\ B-1050 Brussels, Belgium,\\
        Department of Physics, Swansea University, Swansea, SA2 8PP, U.K. \\
        E-mail: \email{D.C.Thompson@Swansea.ac.uk}}

        \abstract{Poisson-Lie (PL) T-duality has received much attention over the last five years in connection with integrable string worldsheet theories.     At the level of the worldsheet, the algebraic structure underpinning these connections is made manifest with the $\mathcal{E}$-model, a first order Hamiltonian description of the string.  The $\mathcal{E}$-model shares many similarities with Double Field Theory (DFT).   We report on recent progress in establishing a precise linkage with DFT as the  target space description of the $\mathcal{E}$-model.    
There are three important outcomes of this endeavor:
          \begin{enumerate}
            \item PL symmetry is made manifest at the level of (generalized) supergravity in DFT.
            \item PL symmetric target spaces are described by a set of generalized frame fields that encode  consistent truncations of supergravity.
            \item PL dualisation rules are made explicit and are readily extended to include the R/R sector of the type II theory.
          \end{enumerate}
These general results are put into context with their application to the the integrable Yang-Baxter model ($\eta$-deformation).  This extended proceedings provides some introductory review of PL  and an orientation to the results of \cite{Hassler:2017yza} and  \cite{Demulder:2018lmj}. }

\FullConference{Corfu Summer Institute 2018 "School and Workshops on Elementary Particle Physics and Gravity"\\
		(CORFU2018)\\
		31 August - 28 September, 2018\\
		Corfu, Greece}

\begin{document}

\section{Introduction}
Abelian T-duality \cite{Buscher:1987sk,Buscher:1987qj} is a fundamental concept in string theory. In its purest form, it states that closed string theory is invariant under the inversion of the radius of a circle in target space. For this duality to be applicable, the target space has to admit one or more abelian isometries. However, in general isometries form non-abelian isometry groups. Hence a natural question is whether there is also a non-abelian T-duality \cite{delaOssa:1992vci} with similar properties. Surprisingly, the answer to this question is harder than initially expected. Trying to extend the worldsheet approach to T-duality from abelian to non-abelian groups, one loses control of both $\alpha'$ and $g_s$ corrections \cite{Giveon:1993ai}. But even at the classical level, where we can neglect these corrections, the dual target space in general lacks the isometries required to recover the original one. Non-abelian T-duality looks at this level like a one-way street \cite{Alvarez:1994dn}.

This puzzle was solved two decades ago in the remarkable work of Klim\v{c}\'ik and \v{S}evera \cite{Klimcik:1995ux,Klimcik:1995dy}. They realized that under special circumstances (dubbed Poisson-Lie symmetry\cite{Klimcik:1995jn}) a target space does not even require isometries to permit T-duality at the classical level. Due to underlying Poisson-Lie group in the construction, this generalized notion of T-duality is called Poisson-Lie T-duality. It contains both the abelian and non-abelian version as special case. Classically Poisson-Lie dual $\sigma$-models are related by a canonical transformation of their phase space and therewith are completely equivalent \cite{Klimcik:1995dy,Sfetsos:1997pi}. As for non-abelian T-duality, the control over quantum correction is in general lost. Missing a full quantum equivalence, one could question the relevance of this duality. On the other hand there are two remarkable properties which still make Poisson-Lie T-duality very useful:
\begin{itemize}
  \item Under mild assumptions, solutions of the one-loop $\beta$-functions are preserved \cite{Sfetsos:1998kr,Sfetsos:2009dj,Valent:2009nv}. Thus, one might suggest that this duality is a map between CFTs. This property makes it a powerful solution generating technique in supergravity (SUGRA). Applications include using non-abelian T-duality to generates new examples of holographic backgrounds \cite{Sfetsos:2010uq,Lozano:2011kb,Lozano:2012au,Lozano:2013oma,Itsios:2013wd}.
  \item All known examples of two dimensional $\sigma$-models which are classically integrable have Poisson-Lie symmetry.
\end{itemize}
The second point is intriguing, because it connects two at first unrelated ideas. Heuristically, a dynamical system is integrable, if its number of degrees of freedom is matched by its independently conserved quantities\footnote{Of course the infinity of degrees of freedom in field theory and even more in a quatum field theory requires some more care.} \cite{Babelon:2003qtg}. Many fundamental systems we encounter in theoretical physics, like the harmonic oscillator have this property. But one should not come to the conclusion that integrability is common. On the contrary it is very rare. The reason why we still come across the few exceptions regularly is because they are exactly solvable and thus a perfect testing ground for ideas and concepts in theoretical physics. An essential example is the classical closed string on AdS$_5\times$S$^5$ \cite{Bena:2003wd}. Due to the AdS/CFT correspondence \cite{Maldacena:1997re} it is conjectured to admit an alternative description in terms of 4D maximal supersymmetric Yang-Mills theory, a conformal field theory (CFT). A challenge in exploring this duality is that always at least one of its sides is strongly coupled and hard to access. Exploiting integrability, there is still a lot of progress recently. For example the anomalous dimensions of certain single trace operators in the CFT can be obtained from integrable spin chains \cite{Minahan:2002ve,Beisert:2003yb}. Instead of going into further details here, we refer to the comprehensive review \cite{Beisert:2010jr}.

To find new integrable two dimensional $\sigma$-models is challenging. A methodical approach is to start with one of the few we know, like the principal chiral model (PCM) \cite{zakharov1978relativistically,Faddeev:1985qu}, and deform it such that classical integrability is preserved \cite{Delduc:2013fga}. An important example for this idea is the Yang-Baxter $\sigma$-model (YB model) \cite{Klimcik:2002zj}, variously known as $\eta$-deformation ($\eta$ is the deformation parameter). It was introduced by Klimcik originally as a model with rich Poisson-Lie T-duality and later discovered to be integrable \cite{Klimcik:2008eq}. More recently, a significant amount of activity was triggered by the extension of the YB model to symmetric spaces \cite{Delduc:2013fga} and eventually the $\eta$-deformation of the full AdS$_5\times$S$^5$ background \cite{Delduc:2013qra}. Another integrable deformation, the $\lambda$-deformation, starts from a Wess-Zumino-Witten (WZW) model \cite{Sfetsos:2013wia}. It seems to be quite different for the $\eta$-deformation, but in fact both are related by Poisson-Lie T-duality and an analytic continuation \cite{Hoare:2015gda,Sfetsos:2015nya,Klimcik:2015gba}. Building on these two, there is now a whole ``zoo'' of integrable deformations. Example are the two, two parameter deformations named biYB model \cite{Klimcik:2014bta,Klimcik:2016rov} and YB Wess-Zumino (WZ) model \cite{Kawaguchi:2011mz,Kawaguchi:2013gma,Delduc:2014uaa,Klimcik:2017ken}. All of them are Poisson-Lie symmetric.

Historically, Poisson-Lie T-duality is approached from the worldsheet perspective. It can be made manifest in a doubled $\sigma$-model, coined the $\mathcal{E}$-model \cite{Klimcik:1995dy,Klimcik:1996nq,Klimcik:2015gba}. Using such a doubled formalism already proved to be convenient for abelian T-duality \cite{Duff:1989tf,Tseytlin:1990va,Tseytlin:1990nb,Hull:2004in,Hull:2006va}. Its two major ingredients are an O($D$,$D$) invariant inner product and a generalized metric. Latter combines the metric and the $B$-field of closed string theory in one object. Half of the coordinates of the doubled target space in this models can be eliminated to recover a standard $\sigma$-model. When this reduction is not unique, it gives rise to T-dual backgrounds. For abelian T-duality there is also an elegant doubled, duality symmetric effective theory on the target space called Double Field Theory (DFT) \cite{Siegel:1993th,Siegel:1993xq,Hull:2009mi,Hohm:2010pp}. For certain calculations it has advantages over the worldsheet treatment. For example the Ramond/Ramond (R/R) sector which is very important to find SUGRA solutions is much easier to access. Thus, one would hope that there is a similar understanding at the level of the target space for Poisson-Lie T-duality. First clues in this direction came from the one-loop $\beta$-functions for the $\mathcal{E}$-model \cite{Avramis:2009xi,Sfetsos:2009vt}. They correspond to the scalar field equations of a gauged supergravity (SUGRA) whose embedding tensor matches the structure coefficients of a doubled Lie group $\mathdsl{D}$. But $\mathdsl{D}$ is nothing else than the $\mathcal{E}$-model's target space. Gauged SUGRAs are further related to DFT by (generalized) Scherk-Schwarz reductions \cite{Scherk:1978ta,Scherk:1979zr,Grana:2012rr,Geissbuhler:2011mx,Aldazabal:2011nj,Catal-Ozer:2017cqr} on group manifolds. Thus what is required is a precise formulation of DFT on group manifolds like $\mathdsl{D}$. It was developed in \cite{Blumenhagen:2014gva,Blumenhagen:2015zma,Bosque:2015jda} and goes under the name \DFTwzw\footnote{This name is a historical artifact, because the first derivation applied closed string field theory to a WZW-model, before the theory was later extended to more general $\sigma$-models. We say more about this point in section~\ref{sec:DFTwzwNSNS}.}.

In this paper, we want to review the connection between \DFTwzw{} and the $\mathcal{E}$-model. Most of the presented results are not new but can be found in the two articles \cite{Hassler:2017yza,Demulder:2018lmj}. Both live on two opposite sides of the spectrum. Because \cite{Hassler:2017yza} is very short, it emphasis the idea which connects \DFTwzw{} with Poisson-Lie symmetry but it does not give many details. On the other hand \cite{Demulder:2018lmj} is rather long and contains a lot of technical parts. They show that all the ideas are on solid grounds but they can also be overwhelming at first. Hence, here we try to rather emphasis conceptual aspects instead of technical ones. While at the same time we provide enough background material to make the beautiful connection between Poisson-Lie T-duality, integrable deformations and DFT hopefully better accessible to members of all three communities. A first impression how fruitful this connection can be is given by the recent activity in this new direction \cite{Lust:2018jsx,Marotta:2018swj,Severa:2018pag,Crow-Watamura:2018liw,Marotta:2019wfq,Sakatani:2019jgu,Catal-Ozer:2019hxw}. But there is still a lot to explore. We address some of the open questions in the conclusions.

\section{$\mathcal{E}$-model and integrable deformations}\label{sec:Emodel}
The coordinate-free approach to general relativity is appreciated for its elegance. It is totally democratic because any choice of coordinates always singles out a particular perspective on a geometry over all others. Eventually, the physics of course does not depend on how we choose coordinates or if we choose any at all. A very similar idea underlies the $\mathcal{E}$-model \cite{Klimcik:1995dy,Klimcik:1996nq,Klimcik:2015gba}. There is only one physical system, given in terms of an action on a two dimensional worldsheet. But depending on how we choose the physical fields, we can interpret it as two or sometimes even more $\sigma$-models on very different looking target spaces. To make this idea more explicit, we start with the ingredients required to construct an $\mathcal{E}$-model:
\begin{itemize}
  \item an even dimensional Lie group $\mathdsl{D}$ equipped with an ad-invariant, non-degenerate, bilinear pairing $\llangle \cdot, \cdot \rrangle$ of split signature on the corresponding Lie algebra $\mathfrak{d}$\label{bullet:LiegroupD}
  \item at least one maximally isotropic subgroup, denoted as $\tilde{H}$
  \item a self-adjoint, involution\footnote{It is possible to drop this condition and study $\mathcal{E}$-models with non-involutive $\mathcal{E}$ \cite{Klimcik:2019kkf}. They give rise to the dressing coset construction \cite{Klimcik:1996np} and capture an even larger class of Poisson-Lie T-duality than the one we discuss here.} $\mathcal{E}: \mathfrak{d} \rightarrow \mathfrak{d}$
\end{itemize}
To capture the dynamics of this model, one employs the Hamiltonian
\begin{equation}
  \mathrm{Ham} = \frac12 \oint d \sigma \llangle j(\sigma), \mathcal{E} j(\sigma) \rrangle
    = \frac12 \oint \dd\sigma\, \llangle \mathdsl{g}^{-1} \partial_\sigma \mathdsl{g}, \mathcal{E} \mathdsl{g}^{-1} \partial_\sigma \mathdsl{g} \rrangle
\end{equation}
with the Lie algebra valued current $j(\sigma) = \mathdsl{g}^{-1} \partial_\sigma \mathdsl{g}$ on $\mathdsl{D}\ni\mathdsl{g}$. To study the dynamics of this Hamiltonian, we further need the Poisson brackets for the current components. To make things a bit less abstract, we choose a basis $\mathdsl{T}_A = \begin{pmatrix}\widetilde T^a & T_a \end{pmatrix}$ where $\widetilde T^a$ are generators of $\tilde{\mathfrak{h}}$, the Lie algebra of $\tilde H$. $T_a$ are the remaining generators, which in general need not form a Lie algebra. In this basis the current components $j_A(\sigma) = \llangle \mathdsl{T}_A , j(\sigma) \rrangle$ are governed by the Poisson brackets
\begin{equation}
  \{ j_A(\sigma), j_B(\sigma') \} = {\mathdsl F}_{AB}{}^C j_C(\sigma) \delta( \sigma - \sigma' ) +
    \eta_{AB} \delta^\prime( \sigma - \sigma' ) \,.
\end{equation} 
This relation relies on two tensors, the structure coefficients $\mathdsl{F}_{AB}{}^C$ of $\mathfrak{d}$ and the pairing. Without loss of generality one can choose the basis $\mathdsl{T}_A$ of the Lie algebra such that
\begin{equation}\label{eqn:pairing}
  [ \mathdsl{T}_A, \mathdsl{T}_B ] = \mathdsl{F}_{AB}{}^C \mathdsl{T}_C 
    \qquad \text{and} \qquad
  \eta_{AB} = \llangle \mathdsl{T}_A, \mathdsl{T}_B \rrangle = \begin{pmatrix} 0 & \delta^a{}_b \\
    \delta_a{}^b & 0
  \end{pmatrix}
\end{equation}
holds.

How is this description connected to a closed string $\sigma$-model, written in terms of a target space metric and $B$-fields? To answer this question, we first switch from the Hamiltonian description to the Lagrangian picture with the action \cite{Klimcik:1995dy}
\begin{equation}\label{eqn:SEmodel}
  S = \frac12 \int \dd\sigma\,\dd\tau \llangle \mathdsl{g}^{-1} \partial_\sigma \mathdsl{g}, \mathdsl{g}^{-1} \partial_\tau \mathdsl{g} \rrangle + \frac1{12} \int \llangle [ \mathdsl{g}^{-1} \dd \mathdsl{g} , \mathdsl{g}^{-1} \dd \mathdsl{g} ], \mathdsl{g}^{-1} \dd \mathdsl{g} \rrangle
  - \int \dd \tau \, \mathrm{Ham}
\end{equation}
where the first two terms on the right hand side govern a standard WZW-model on $\mathdsl{D}$. We further split the Lie group element 
\begin{equation}\label{eqn:splitg}
  \mathdsl{g} = \tilde{h} m \qquad \tilde{h} \in \tilde{H}
\end{equation}
into a coset representative $m$ and an element of the maximally isotropic subgroup. Because of the special properties of the subgroup $\tilde{H}$, $\tilde{h}$ only appears in the combination $\tilde{h}^{-1} \partial_\sigma \tilde{h}$ in the field equations and we are able to integrate it out. After this step the remaining target space is $M$=$\tilde{H}\backslash\mathdsl{D}$ and the action becomes \cite{Klimcik:1996nq,Klimcik:2015gba,Klimcik:2019kkf}
\begin{equation}\label{eqn:SHt/D}
  S = \frac14 \int \dd \xi^+ \, \dd \xi^- \llangle \left( 1 - 2 \mathcal{P} (\mathcal{E}) \right) m^{-1} \partial_+ m, m^{-1} \partial_- m \rrangle + \frac1{12} \int \llangle [ m^{-1} \dd m, m^{-1} \dd m ], m^{-1} \dd m \rrangle\,.
\end{equation}
Here we also switched to light cone coordinates $\xi^\pm = 1/2 (\tau \pm \sigma)$. $\mathcal{P}(\mathcal{E})$ is a projector that originates from integrating out the $\tilde{h}$ contributions. It is defined by its image and kernel
\begin{equation}
  \textrm{Im}\, {\mathcal P} = \tilde{\mathfrak{h}} \,, \quad  \textrm{Ker}\, {\mathcal P} = (1 +  \textrm{ad}_{m^{-1}} \cdot   {\mathcal E} \cdot  \textrm{ad}_m ) \mathfrak{d} \, .
\end{equation}
From the first terms in \eqref{eqn:SHt/D}, we can read off the target space metric (symmetric part) and the $B$-field (anti-symmetric part). The second term is a WZ-term. It captures an $H$-flux which cannot written as $H=\dd B$ such that $B$ is globally defined on $M$. There are three important lessons which can be learned from this short derivation:
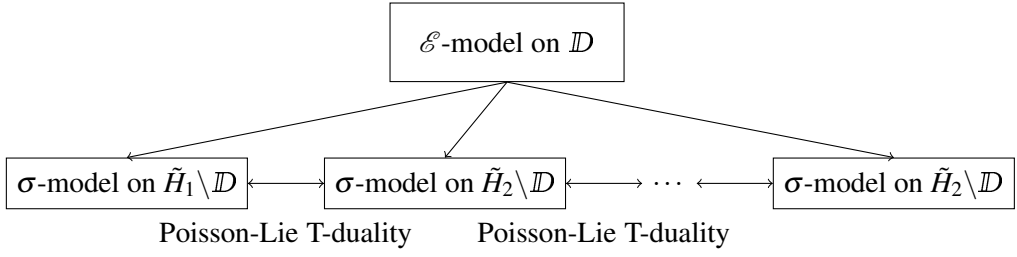
\begin{figure}[!t]
  \centering
  \begin{tikzpicture}
    \node[rectangle,draw,name=Emodel,inner sep=1em]{$\mathcal{E}$-model on $\mathdsl{D}$};
    \node[rectangle,draw,name=H1,at={(Emodel.south)},anchor=north,
      xshift=-5cm,yshift=-1cm]{$\sigma$-model on $\tilde{H}_1 \backslash \mathdsl{D}$};
    \node[rectangle,draw,name=H2,at={(H1.east)},anchor=west,
      xshift=1cm]{$\sigma$-model on $\tilde{H}_2 \backslash \mathdsl{D}$};
    \node[name=dots,at={(H2.east)},anchor=west,
      xshift=1cm]{$\dots$};
    \node[draw,rectangle,name=Hn,at={(dots.east)},anchor=west,
      xshift=1cm]{$\sigma$-model on $\tilde{H}_2 \backslash \mathdsl{D}$};
    \draw[->] (Emodel.south) -- (H1.north);
    \draw[->] (Emodel.south) -- (H2.north);
    \draw[->] (Emodel.south) -- (Hn.north);
    \draw[<->] (H1.east) -- (H2.west) node[midway,below,yshift=-1em] {Poisson-Lie T-duality};
    \draw[<->] (H2.east) -- (dots.west) node[midway,below,yshift=-1em] {Poisson-Lie T-duality};
    \draw[<->] (dots.east) -- (Hn.west);
  \end{tikzpicture}
  \caption{If the Lie group $\mathdsl{D}$ of the $\mathcal{E}$-models admits $n>1$ maximally isotropic subgroups, it gives rise to $n$ $\sigma$-models on in general very different looking target spaces. But because all of them originate from the same $\mathcal{E}$-model they are classically still absolutely equivalent. Poisson-Lie T-duality (sometimes also called plurality if $n>2$) is the involved map connecting these target spaces.}
  \label{fig:PLTDsigmamodels}
\end{figure}
\begin{itemize}
  \item If there is more than one maximally isotropic subgroup, like depicted in figure~\ref{fig:PLTDsigmamodels}, we can derive for each a $\sigma$-model. These models describe very different target spaces but are equivalent because they originate from the same $\mathcal{E}$-model. They are Poisson-Lie T-dual. 
  \item In practice the projector $\mathcal{P}$ quickly gets complicated if we choose explicit coordinates on $M$. Knowing just the metric and $B$-field it is hard to even figure out whether the $\sigma$-model originated from an $\mathcal{E}$-model or not. Thus, Poisson-Lie T-duality is a hidden symmetry at this level and only is manifest in the $\mathcal{E}$-model formulation.
  \item\label{bullet:genEmodel} There is a slight generalization of the original $\mathcal{E}$-model which is important for the DFT description. Originally, the operator $\mathcal{E}$ is constant on $\mathdsl{D}$. But to go from \eqref{eqn:SEmodel} to \eqref{eqn:SHt/D} it is sufficient to only require that it is independent of $\tilde H$. We will come back to this point in much more detail in section~\ref{sec:PLsymts}.
\end{itemize}

Why is this approach so well suited to study integrable deformations? Assume we want to study an integrable $\sigma$-model on a group manifold $G$. Our first task is to understand how we can find all independently conserved charges integrability requires. The $\mathcal{E}$-model makes this step easy. First, we identify $\mathdsl{D}$ with the complexification $G^{\mathbb{C}}$ of a $D$-dimensional Lie group $G$ and study the time evolution of the current $j(\sigma)$. Using the Poisson bracket and the Hamiltonian, we find
\begin{equation}\label{eqn:ddtj}
  \partial_\tau j(\sigma) = \{\mathrm{Ham}, j(\sigma)\} = \partial_\sigma ( \mathcal{E} j(\sigma) ) + [ \mathcal{E} j(\sigma) , j(\sigma) ]\,. 
\end{equation}
Because $\mathdsl{D}=G^{\mathbb{C}}$, the Lie algebra $\mathfrak{d}$ is complex and has a $\mathfrak{g}$-valued real and imaginary part. Hence, we decompose the current $j(\sigma) = \mathcal{R}(\sigma) + i \mathcal{J}(\sigma)$ accordingly. $\mathcal{E}$ acts by swapping $\mathcal{R}(\sigma)$ and $\mathcal{I}(\sigma)$. A short calculation \cite{Klimcik:2008eq,Klimcik:2017ken} shows that \eqref{eqn:ddtj} equivalently defines a family of  $\mathfrak{g}$-valued, flat\footnote{Flat just means that the field strengths $\dd A(\lambda) + [A(\lambda), A(\lambda)] = 0$ have to vanish for all values of $\lambda \in \mathbb{C} \backslash \{-1,1\}$.} connections
\begin{equation}\label{eqn:laxconnection}
  A(\lambda) = \frac{\mathcal{R} + \mathcal{J}}{1+\lambda} \dd \xi^+ + \frac{\mathcal{R} - \mathcal{J}}{1-\lambda} \dd \xi^-  \qquad \xi^\pm = \frac12 ( \tau \pm \sigma )
\end{equation}
on the worldsheet. $A(\lambda)$ is called Lax connection and ensures that there are infinitely many integrals of motion as required for a integrable system.

The idea behind integrable deformations is to keep the Lax connection \eqref{eqn:laxconnection} but change the decomposition of $j(\sigma)$ into $\mathcal{R}(\sigma)$ and $\mathcal{J}(\sigma)$. We cannot change it arbitrarily, but only such that the flatness of $A(\lambda)$ still captures the time evolution of $j(\sigma)$. An explicit example is the Yang-Baxter deformation. Normally, its deformation parameter is denotes as $\eta$. To avoid a conflict with the tensor $\eta_{AB}$ in \eqref{eqn:pairing}, we call it $\kappa$ here. The canonical basis $\mathdsl{T}_A=(\tilde T^a, T_a)$ for the Lie algebra $\mathfrak{d}$ is chosen such that the generators $T_a$ generate $G$. Furthermore, we choose $\tilde T^a$ such that we obtain the commutators
\begin{equation}\label{eqn:basisdeta}
  [T_a , T_b] = f_{ab}{}^c T_c \,, \qquad
  [T_a , \tilde T^b] = 2 R^{[b d} f_{ad}{}^{c]} T_c - f_{ac}{}^b \tilde T^c \,, \qquad
  [\tilde T^a, \tilde T^b] = 2 R^{[a d} f_{cd}{}^{b]} \tilde T^c\,.
\end{equation}
Beside the structure coefficients $f_{ab}{}^c$ of $\mathfrak{g}$, the $R$-matrix appears here. It plays a pivotal role for integrable deformations. The map $R: \mathfrak{g} \rightarrow \mathfrak{g}$ is a solution of the modified classical Yang-Baxter equation mCYBE
\begin{equation}\label{eq:YBE}
  [R X, R Y]- R \left([R X, Y]  +[X,  R Y] \right) = [X, Y]\, , \quad \forall X,Y \in \mathfrak{g}  \,. 
\end{equation}
It immediately follows from the Jacobi identity for the subgroup generators $\tilde T^a$. Writing it in our basis $T_a \rightarrow R_a{}^b T_b$ implies the index structure of $R$. To obtain $R^{ab}$, which is skew-symmetric, we have to raise the first index with the inverse Killing metric $k^{ab}$ of $\mathfrak{g}$. Finally, we have to relate the current components
\begin{equation}
  \mathcal{R}_a(\sigma) = \frac1{\sqrt{\kappa}} j_a(\sigma) \qquad
  \mathcal{J}_a(\sigma) = \sqrt{\kappa} \left( k_{ab} \tilde j^b(\sigma) - R_a{}^b j_b(\sigma) \right)\,.
\end{equation}
For $\kappa$=1, we recover the starting point $j(\sigma) = \mathcal{R}(\sigma) + i \mathcal{J}(\sigma)$ of our construction. If $\kappa$ vanishes, the Lie group $\mathdsl{D}$ undergoes a group contraction and becomes $T^* G$ instead of $G^{\mathbb{C}}$. The resulting $\sigma$-model is the principal chiral model on $G$. Hence, $\kappa$ measures the deviation from the PCM.

This is just one example of an integrable deformation of the PCM. There are also multiparameter deformations like the biYB model \cite{Klimcik:2014bta,Klimcik:2016rov} or the YB WZ model \cite{Delduc:2014uaa,Klimcik:2017ken}. Their construction involves more algebra but follows the same idea. Instead of deforming the PCM one could also start with a WZW-model to obtain the $\lambda$-deformation \cite{Sfetsos:2013wia}. However, all these deformations have one thing in common: Their integrability is quite opaque in the $\sigma$-model \eqref{eqn:SHt/D} but easy to spot in the $\mathcal{E}$-model. This is remarkable because $\mathcal{E}$-models were introduced to study Poisson-Lie T-duality and a priori do not know anything about integrability.

\section{\DFTwzw{}, low energy effective target space theory of the $\mathcal{E}$-model}\label{sec:DFTwzwNSNS}
String theory can be approached from two different perspectives. On the worldsheet we are focused on two-dimensional $\sigma$-models. While from the target space point of view one is more interested in low energy effective theories, like SUGRA. Both have their strengths and of course they are related. Ideally one seeks for a good understanding of both to use the best of both worlds.

We reviewed the worldsheet approach to the $\mathcal{E}$-model and its applications in the last section. What is the corresponding target space story? A quick answer is that the $\mathcal{E}$-model is equivalent to the standard $\sigma$-model \eqref{eqn:SHt/D} on the target space $M=\tilde{H}\backslash\mathdsl{D}$. Thus, we know that its low energy effective theory has to be the NS/NS (NS=Neveu–Schwarz) sector of SUGRA\footnote{Here we assume that $\tilde H$ is unimodular. If we drop this assumption, which is required for the $\eta$-deformation, we have to consider generalized SUGRA \cite{Arutyunov:2015mqj}. We say more about this subtlety later.}. The problem with this answer is that the nice structure of the $\mathcal{E}$-model is lost or at least is very opaque. It would be much better to find an effective theory with makes directly contact with the Lie group $\mathdsl{D}$, $\mathcal{E}$ and the two tensors $\mathdsl{F}_{AB}{}^C$ and $\eta_{AB}$. At this point one could start to guess that this theory is connected to DFT \cite{Siegel:1993th,Siegel:1993xq,Hull:2009mi,Hohm:2010pp}. There is a ``doubled'' target space $\mathdsl{D}$ and an O($D$,$D$) invariant metric $\eta_{AB}$. We also can identify $\mathcal{E}$ with
\begin{equation}
  \mathcal{H}_{AB} = \llangle \mathdsl{T}_A , \mathcal{E} \mathdsl{T}_B \rrangle\,.
\end{equation}
Because $\mathcal{E}$ is self-adjoint, $\mathcal{H}_{AB}$ is symmetric. Furthermore, because $\mathcal{E}$ is involutive it satisfies $\mathcal{H}_{AC} \eta^{CD} \mathcal{H}_{DB} = \eta_{AB}$, where $\eta^{AB}$ is the inverse of $\eta_{AB}$. In DFT, one calls this tensor generalized metric. After all this evidence DFT seems to be the correct framework to describe the low energy effective target space theory of the $\mathcal{E}$-model. However, there are some subtleties which we have to address.

\subsection{Action and symmetries}
One conceptual challenge in DFT is the interpretation of the doubled space. For strings on a torus, the torus coordinates $x^i$ are conjugate to the center of mass momentum of the string. But because closed strings can also wind around non-contractible cycles of the tours, they in general have winding modes. For them additional conjugate ``winding'' coordinates $\tilde x_i$ are introduced and the doubled space of DFT is born. But intriguingly, DFT proved useful for backgrounds without any non-contractible cycles like spheres, too. Prominent examples are generalized Scherk-Schwarz reductions \cite{Grana:2012rr,Geissbuhler:2011mx,Aldazabal:2011nj,Catal-Ozer:2017cqr} which give rise to gauged supergravities. To approach this puzzle the derivation of DFT from closed string field theory (CSFT) on a torus \cite{Hull:2009mi} was repeated in \cite{Blumenhagen:2014gva} for WZW-models. These $\sigma$-models include target spaces without non-contractible cycles. A simple example is the three sphere with $H$-flux, captured by the SU(2) WZW-model. Furthermore, they have a nice CFT description which is required to apply CSFT. For obvious reasons the resulting theory was dubbed \DFTwzw{}. Its action \cite{Blumenhagen:2015zma} 
\begin{equation}\label{eqn:Sdftwzw}
\begin{aligned}
  S_{\mathrm{NS}} =& \int d^{2D} X e^{-2d} \Big(  \frac{1}{8} \mathcal{H}^{CD} \nabla_C \mathcal{H}_{AB}
    \nabla_D \mathcal{H}^{AB} -\frac{1}{2} \mathcal{H}^{AB} \nabla_{B} \mathcal{H}^{CD}
    \nabla_D \mathcal{H}_{AC} \\
    & - 2 \nabla_A d \nabla_B \mathcal{H}^{AB} + 4 \mathcal{H}^{AB} \nabla_A d \nabla_B d + 
    \frac{1}{6} \mathdsl{F}_{ACD} \mathdsl{F}_B{}^{CD} \mathcal{H}^{AB} \Big)\,,
  \end{aligned} 
\end{equation}
where indices are lowered/raised with $\eta_{AB}$/$\eta^{AB}$, looks very similar to the original Hohm-Hull-Zwiebach (HHZ) action \cite{Hohm:2010pp}. The most important differences are:
\begin{itemize}
  \item Instead of using partial derivatives, this action employs covariant derivatives $\nabla_A$.
  \item There is an additional term in the second line which involves the structure coefficients $\mathdsl{F}_{AB}{}^C$ of the Lie algebra $\mathfrak{d}$.
  \item The doubled space does not require the notion of winding coordinates anymore. It is just the Lie group $\mathdsl{D}$\footnote{
    For a WZW-model on the group manifold $G$, the doubled space $\mathdsl{D}=G\times G$ arises \cite{Blumenhagen:2014gva}. The two factors originate from the left- and right-moving sector of the closed string. But remarkably this restriction on $\mathdsl{D}$ can be dropped without any problems \cite{Blumenhagen:2015zma}. $\mathdsl{D}$ just has to be an even dimensional Lie group whose Lie algebra is equipped with an ad-invariant, non-degenerate, bilinear pairing of split signature. From this point of view the name \DFTwzw{} should be abandoned in favor of Double Field Theory on group manifolds. But because it is so short it still survived till today.} that we introduced for the $\mathcal{E}$ in this first bullet point on page~\pageref{bullet:LiegroupD}.
\end{itemize}
The generalized dilaton $d$ is a new scalar that we did not encounter in the discussion of the $\mathcal{E}$-model. It describes the dilaton which on the worldsheet only becomes relevant in one-loop effects. This is the first advantage of the target space description we encounter: The dilaton is automatically taken care of and does not require further work.

Compared to HHZ DFT most striking is probably the appearance of a covariant derivative. It is covariant with respect to conventional $2D$-diffeomorphisms on the Lie group $\mathdsl{D}$. Furthermore it is compatible with $\eta_{AB}$ ($\nabla_A \eta_{BC} = 0$). In general there is a one parameter family of $\eta_{AB}$-compatible covariant derivatives $\nabla_A^{(t)}$ on $\mathdsl{D}$. They act on an arbitrary vector $V^B$ of weight $w(V^B)$ by
\begin{equation}
  \nabla^{(t)}_A V^B = \mathdsl{E}_A{}^I \partial_I V^B + t \mathdsl{F}_{AC}{}^B V^C - w(V^B) \mathdsl{E}_A{}^I \partial_I \log | \det \mathdsl{E} | V^B
\end{equation}
where $\mathdsl{E}_A{}^I$ denotes the vector fields on $\mathdsl{D}$ which generate right translation. They arise as the inverse transpose of $\mathdsl{E}^A{}_I$ (in a determinate just $\mathdsl{E}$) which captures the left-invariant Maurer-Cartan form $\mathdsl{T}_A \mathdsl{E}^A{}_I d X^I = \mathdsl{g}^{-1} \dd \mathdsl{g}$. Curvature and torsion for $\nabla_A^{(t)}$ are defined by the commutator
\begin{equation}
  [\nabla_A^{(t)}, \nabla_B^{(t)}] V_C = R_{ABC}{}^D V_D - T_{AB}{}^D \nabla_D^{(t)} V_C
\end{equation}
and read
\begin{equation}
  R_{ABC}{}^D = (t - t^2) \mathdsl{F}_{AB}{}^E \mathdsl{F}_{EC}{}^D \qquad
  T_{AB}{}^C = (2 t - 1) \mathdsl{F}_{AB}{}^C \,.
\end{equation}
For $t$=1/2 we obtain the torsion-free Levi-Civita connection. Connections without curvature arise for $t$=0 and $t$=1. They correspond to the left- and right-trivialization of the tangent bundle $T \mathdsl{D}$. But remarkably $\nabla_A$ in the action \eqref{eqn:Sdftwzw} is none of them. Instead the explicit CSFT calculations require
\begin{equation}
  \nabla_A := \nabla_A^{(1/3)}\,.
\end{equation}
Another covariant derivative which we will need from time to time is the flat derivative
\begin{equation}
  D_A := \nabla_A^{(0)}\,.
\end{equation}

Once we have an action, we should ask about its symmetries. Here we present them acting on a vector density or $(1,0)$ tensor density. You might wonder why we do this because neither of the fundamental field $\mathcal{H}_{AB}$ nor $d$ is a vector density. However the transformation of arbitrary tensors can easily be extracted from the transformation of $(1,0)$ tensor density by imposing the Leibniz rule and taking into account that $\eta_{AB}$ is always invariant.

A {\it global O($D$,$D$) transformations} symmetry arises from the fact that we can freely choose the basis for our Lie algebra as long as we do not spoil the canonical form of $\eta$ in \eqref{eqn:pairing}. It is mediated by 
\begin{equation}
  V^A \rightarrow \mathcal{O}^A{}_B V^B 
  \qquad \text{with} \qquad
  \mathcal{O}^A{}_C \mathcal{O}^B{}_D \eta^{CD} = \eta^{AB} \,.
\end{equation}
There are also two local symmetries. Their infinitesimal version changes the coordinates on $\mathdsl{D}$
\begin{equation}
  X^I \rightarrow X^I + \xi^A \mathdsl{E}_A{}^I
\end{equation}
and at the same time applies
\begin{equation}
  V^A \rightarrow V^A + \delta V^A
\end{equation}
to a vector density. For the $\delta V^A$ in this equation we have two choices
\begin{itemize}
  \item {\it Conventional $2D$-diffeomorphisms} mediated by the Lie derivative
    \begin{equation}\label{eq:2Ddiffeo}
      \delta V^A = L_\xi V^A = \xi^BD_B V^A  + w(V^A) D_B \xi^B V^A \,.
    \end{equation}
    You might notice that this is not the Lie derivative you would expect for a vector density. It is the one for a scalar density instead. The key to this puzzle is to remember that the Lie derivative acts on coordinate (curved) indices $I,J,\dots,Z$. For example $\mathdsl{E}_A{}^I$ transforms as a vector with the familiar expression
    \begin{equation}
      L_{\mathdsl{E}_A} \mathdsl{E}_B{}^I = \mathdsl{E}_A{}^J \partial_J \mathdsl{E}_B{}^I - \mathdsl{E}_B{}^J \partial_J \mathdsl{E}_A{}^I
    \end{equation}
    for the Lie derivative. But $V_A$ does not carry any curved indices. It arises from the contraction $V^A$=$\mathdsl{E}^A{}_I V^I$, where $\mathdsl{E}^A{}_I$ is a one form and $V^I$ is a vector density. By applying the Leibniz rule we easily verify that $V^A$ has to transform as a scalar density. The same argument holds for arbitrary tensors with just flat indices $A,B,\dots,H$. This symmetry is manifest, because the action is written in terms of conventional covariant derivatives.
  \item {\it Generalized diffeomorphisms} mediated by the generalized Lie derivative
  \begin{equation}
    \delta V^A = {\cal L}_\xi V^A = \xi^B \nabla_B V^A -  V^B \nabla_B \xi^A + \eta^{AB}\eta_{CD}  V^C  \nabla_B \xi^D  + w(V^A) \nabla_B \xi^B V^A\,.
  \end{equation}
  This symmetry is not manifest because the covariant derivatives in the action are not covariant with respect to generalized diffeomorphisms.  
\end{itemize}
Taking a look at table~\ref{tab:trans}, we find  how the objects we encounter in the action \eqref{eqn:Sdftwzw} transform under these three symmetries. With this information it is very easy to verify that the action is invariant under global O($D$,$D$) transformations and conventional 2$D$-diffeomorphisms. For generalized diffeomorphisms we have to work more because they are not manifest. But in the end we come to the same conclusion modulo a small subtlety;
\begin{table}[!t]
\begin{center}
   \begin{tabular}{r|lll}
          object & gen.-diffeomorphisms & $2D$-diffeomorphisms &
            global $O(D,D)$\\\hline
          $\mathcal{H}_{AB}$ & $(0,2)$ tensor  & scalar & $(0,2)$ tensor \\
          $\nabla_A d$ & not covariant & scalar & $(0,1)$ tensor \\
          $e^{-2d}$    & scalar density, $w(e^{-2d})$=$1$ & 
                         scalar density, $w(e^{-2d})$=$1$ & invariant   \\
          $\eta_{AB}$  & invariant & invariant & invariant  \\
          $\mathdsl{F}_{AB}{}^C$ & invariant & invariant & tensor \\
          $\mathdsl{E}_A{}^I$    & invariant     & vector & $(0,1)$ tensor \\
          $D_A$        & not covariant & covariant & covariant  \\
          $\nabla_A$   & not covariant & covariant & covariant  \\
          $S_{\mathrm{NS}}$ & invariant& invariant & invariant  \\
        \end{tabular}
       \caption{Transformation properties of objects under \DFTwzw{} symmetries.}
       \label{tab:trans}
 \end{center}      
 \end{table}%
while conventional $2D$-diffeomorphisms are valid in general, generalized diffeomorphisms need an additional constraint to close and leave the action invariant. It is called the strong constraint or the section condition and requires that
\begin{equation}\label{eqn:SC}
  D_A \, \cdot \, D^A \, \cdot = 0
\end{equation}
annihilates arbitrary combinations of fields and gauge transformation parameter, denoted here by $\cdot$\,. In closed string field theory this constraint originates from the requirement that in a three point interaction the incoming states but also to outgoing state are level matched. From a conceptual point of view it tells us that the full group manifold $\mathdsl{D}$ is not our physical target space $M$. $M$ is embedded into $\mathdsl{D}$ such that $\dim M$=$1/2\dim\mathdsl{D}$=$D$ and the section condition controls this embedding. We already followed a very similar argument for the $\mathcal{E}$-model. In the last bullet point on page~\pageref{bullet:genEmodel}, $\mathdsl{E}$ is required to be independent of the maximally isotropic subgroup $\tilde H$. Thus, it is not hard to conjecture that solutions of \eqref{eqn:SC} are somehow connected to maximally isotropic subgroups of $\mathdsl{D}$. The next subsection will develop this idea.

\subsection{Section condition and maximally isotropic subgroup}\label{sec:solSC}
The connection between \DFTwzw{} and the $\mathcal{E}$-model took some time to be elucidated. The theory introduced in the last subsection has been around for more than two years before it was related to Poisson-Lie T-duality\cite{Hassler:2017yza}. The crucial insight was to understand how to solve the section condition \eqref{eqn:SC} in the most general way \cite{Hassler:2016srl}. There is a trivial solution where all the fields are constant and we will see in section~\ref{sec:PLsymts} that it is special. But one would expect to find a much larger class of solutions. In order to construct them in a methodical way, it is instructive to rewrite the section condition in terms of partial derivatives
\begin{equation}
  \eta^{IJ} \partial_I \,\cdot\, \partial_J \,\cdot\, = 0
  \qquad
  \eta^{IJ} = \mathdsl{E}_A{}^I \eta^{AB} \mathdsl{E}_B{}^J\,.
\end{equation}
In the HHZ DFT, $\eta^{IJ}$ would be constant and of canonical form. Here it is not, because in general $\mathdsl{E}_A{}^I$ is not an O($D$,$D$) element. The particular $\eta^{IJ}$ we obtain depends on the coordinates we choose on $\mathdsl{D}$. To be explicit, we split these coordinates $X^I = ( x^i \,\,\, \tilde x^{\tilde i} )$ into a physical ($x^i$) and ``unphysical'' ($\tilde x^{\tilde i}$) part. Both $i$ and $\tilde i$ run from one to $D$. If all our fields and gauge transformation parameters just depend on $x^i$ the section condition is solved if we can find a parameterization of $\mathdsl{D}$ such that
\begin{equation}
  \eta^{IJ} = \begin{pmatrix}
    \eta^{ij} & \eta^{i\tilde j} \\
    \eta^{\tilde ij} & \eta^{\tilde i\tilde j}
  \end{pmatrix} =
  \begin{pmatrix}
    0 & \bullet \\
    \bullet & \bullet
  \end{pmatrix}
\end{equation}
where $\bullet$ is just a place holder from some arbitrary components we do not care about. Inverting this matrix, we find
\begin{equation}\label{eqn:etaIJ}
  \eta_{IJ} = \begin{pmatrix}
    \eta_{ij} & \eta_{i\tilde j} \\
    \eta_{\tilde i j} & \eta_{\tilde i\tilde j}
  \end{pmatrix} =
  \begin{pmatrix}
    \bullet & \bullet \\
    \bullet & 0
  \end{pmatrix}\,.
\end{equation}
This tells us to find a solution of the section condition is equivalent to choose the group element $\mathdsl{g}\in\mathdsl{D}$ such that \cite{Hassler:2016srl}
\begin{equation}
  \eta_{\tilde i\tilde j} = \llangle \mathdsl{g}^{-1} \tilde \partial_{\tilde i} \mathdsl{g} , \mathdsl{g}^{-1} \tilde \partial_{\tilde i} \mathdsl{g} \rrangle = 0\,.
\end{equation}
Employing the same splitting for $\mathdsl{g}$ as for the $\mathcal{E}$-model in \eqref{eqn:splitg}
\begin{equation}
  \mathdsl{g}(X^I) = \tilde h(\tilde x^{\tilde i}) m(x^i)
    \qquad \tilde h \in \tilde H
\end{equation}
but written explicitly in our coordinate system on $\mathdsl{D}$, this equation simplifies to
\begin{equation}
  \eta_{\tilde i\tilde j} = \llangle \tilde h^{-1} \tilde\partial_{\tilde i} \tilde h , 
    \tilde h^{-1} \tilde\partial_{\tilde j} \tilde h \rrangle = 0\,.
\end{equation}
It shows that each maximally isotropic subgroup of $\mathdsl{D}$ gives rise to a solution of the section condition. This is the final ingredient to make full contact with the $\mathcal{E}$-model.

Fixing $\eta_{IJ}$ to the particular form \eqref{eqn:etaIJ} partially breaks conventional 2$D$-diffeomorphisms. Arbitrary coordinate transformations would give non-vanishing contributions to the $0$ block in the lower right corner and therefore spoil the section condition solution. This is good news, because the $\sigma$-model on the physical space has just $D$-diffeomorphisms and $B$-field gauge transformations as local target space symmetries. They are mediated by generalized diffeomorphisms. But even if we partially break conventional 2$D$-diffeomorphisms with the section condition, don't we still end up in a situation with local symmetries who have no physical explanation? Luckily, the answer is no because both are not completely independent. All conventional 2$D$-diffeomorphisms which preserve \eqref{eqn:etaIJ} can be written in terms of generalized diffeomorphisms. They do not contribute any additional local symmetries. The world local is very important here. Because under certain conditions, which we explore in section~\ref{sec:PLsymts}, very few conventional 2$D$-diffeomorphisms can survive independently as global symmetries of the theory. They mediate Poisson-Lie T-duality.

\subsection{Equivalence to (generalized) SUGRA}\label{sec:toSUGRANS}
We mentioned in the introduction to this section that the low energy effective action of the $\mathcal{E}$-model should be eventually equivalent to the NS/NS sector of SUGRA (if $\tilde H$ is unimodular). How can we see that \eqref{eqn:Sdftwzw} is just the SUGRA action in disguise. The key to make this connection is called generalized frame field and denoted by $\widehat{E}_A{}^{\hat I}$.

Here a new kind of index appears and to see how it fits into the story lets quickly recap what indices we have already encountered. We started with $A,B,\dots$ in our discussion of the $\mathcal{E}$-model. They capture the adjoint representation of $\mathdsl{D}$'s Lie algebra. Moreover, we have at least one maximally isotropic subgroup $\tilde{H}\subset\mathdsl{D}$. Its generators are labeled by $a,b,\dots$ running from one to $D$. Eventually we introduce coordinates $X^I$ on $\mathdsl{D}$ (at least locally). They carry indices $I,J,\dots$ with are further split into $x^i$ for physical directions or $\tilde x^{\tilde i}$ for unphysical ones. This convention is quite different from generalized geometry \cite{Hitchin:2004ut,Gualtieri:2003dx} which would allow to combine the SUGRA metric and $B$-field into a generalized metric. But of course we found already a lot of evidence that they have to be related. The generalized frame field $\widehat{E}_A{}^{\hat I}$ is the bridge between them. It carries the index $A$, which is the natural choice in the context of the $\mathcal{E}$-model, and $\hat{I}$, the canonical index of generalized geometry. A vector carrying $\hat I$ has the decomposition $\widehat{V}^{\hat I} = ( V_i \,\,\, V^i )$ where $i$ runs from 1 to $D$. To lower/raise these indices, we apply
\begin{equation}\label{eqn:etaGG}
  \widehat{\eta}_{\hat I\hat J} = \begin{pmatrix} 0 & \delta^i{}_j \\
    \delta_i{}^j & 0
  \end{pmatrix}
  \qquad \text{or} \qquad
  \widehat{\eta}^{\hat I\hat J} = \begin{pmatrix} 0 & \delta_i{}^j \\
    \delta^i{}_j & 0
  \end{pmatrix}\,.
\end{equation}
Thus these indices are indeed the ones we know from generalized geometry on the generalized tangent space $T M \oplus T^* M$ of the physical target space $M = \tilde H \backslash \mathdsl{D}$. As an additional safety measure we add a bit of redundancy by decorating quantities with ``hated indices'' with an additional hat. Therefore instead of $V^{\hat I}$ we rather prefer $\widehat{V}^{\hat I}$. Doing so reduces the danger of causing a lot of confusion by accidentally dropping a small hat on an index. Figure~\ref{fig:indices} summarizes our index convention.
\begin{figure}[!b]
  \centering
  \begin{tabular}{rclrcl|rcl}
      \multicolumn{6}{c|}{\DFTwzw{} / $\mathcal{E}$-model} &
      \multicolumn{3}{c}{generalized geometry} \\[0.75em]
      \multicolumn{3}{c}{curved indices $I, J, \dots , Z$} & 
      \multicolumn{3}{c|}{flat indices $A, B, \dots , H$} &  
      \multicolumn{3}{c}{$\hat{I}, \hat{J}, \dots, \hat{Z}$} \\[0.5em]
      $\eta_{IJ}$&=&$\begin{pmatrix} \bullet & 0 \\ \bullet & \bullet \end{pmatrix}$ \quad 
      \eqref{eqn:etaIJ} &
      $\eta_{AB}$&=&$\begin{pmatrix} 0 & \delta^a{}_b \\\delta_a{}^b & 0 \end{pmatrix}$ \quad
      \eqref{eqn:pairing} & 
      $\widehat{\eta}_{\hat I\hat J}$&=&$\begin{pmatrix} 0 & \delta^i{}_j \\ \delta_i{}^j & 0
      \end{pmatrix}$ \quad
      \eqref{eqn:etaGG} \\[1.5em]
      $\tikzmarkMath{VI}{V^I}$&=&$\begin{pmatrix} V^i & V^{\tilde i} \end{pmatrix}$ &
      $\tikzmarkMath{VA}{V^A}$&=&$\begin{pmatrix} V_a & V^a \end{pmatrix}$ &
      $\tikzmarkMath{VhatI}{\widehat{V}^{\hat I}}$ & = & $\begin{pmatrix} V_i & V^i \end{pmatrix}$ 
  \end{tabular}%
  \vspace{2em}%
  \tikz[overlay,remember picture]{
    \draw[<-] (VI) to[bend right=25] node[midway,anchor=north] {$\mathdsl{E}_A{}^I$} (VA);
    \draw[->] (VA) to[bend right=25] node[midway,anchor=north] {$\widehat{E}_A{}^{\hat I}$} (VhatI);
  }%
  \caption{Summary of the different indices we use in this paper. Uppercase indices always run from one to $2D$. They always decompose into to lowercase indices which run only from one to $D$.}
  \label{fig:indices}
\end{figure}

Because we are not concerned with global issues at the moment, for us $\widehat{E}_A{}^{\hat I}$ is just a $2D\times 2D$ matrix. But it is not arbitrary. It has to satisfy the three defining properties:
\begin{itemize}
  \item $\widehat{E}_A{}^{\hat I}$ is an O($D$,$D$) element
    \begin{equation}
      \widehat{E}_A{}^{\hat I} \eta^{AB} \widehat{E}_B{}^{\hat J} = \widehat{\eta}^{\hat I\hat J}\,.
    \end{equation}
  \item $\widehat{E}_A{}^{\hat I}$ only depends on the physical coordinates $x^i$ we defined in the last subsection.
  \item $\widehat{E}_A{}^{\hat I}$ satisfies the frame algebra
    \begin{equation}
      \widehat{{\cal L}}_{\widehat{E}_A} \widehat{E}_B{}^{\hat I} = \mathdsl F_{AB}{}^C \widehat{E}_C{}^{\hat I}
    \end{equation}
    where $\widehat{{\cal L}}$ is the  generalised Lie derivative of generalised geometry
    \begin{equation}
      \widehat{{\cal L}}_{\widehat{\xi}} \widehat{V}^{\hat I} = \widehat{\xi}^{\hat J} \partial_{\hat J} \widehat{V}^{\hat I} + ( \partial^{\hat I} \widehat{\xi}_{\hat J} - \partial_{\hat J} \widehat{\xi}^{\hat I} ) \widehat{V}^{\hat J}
    \end{equation}
    and $\partial_{\widehat{I}} = ( 0 \,\,\, \partial / \partial x^i )$.
\end{itemize}
Such generalized frame fields are also crucial for generalized Scherk-Schwarz reductions. Thus, their construction is interesting on its own right. Explicit examples for $D$=3 are presented in \cite{Dibitetto:2012rk}. We cope with this point later. For the moment, assume that we have a generalized frame field which satisfies these three conditions. It allows us to make the following identifications
\begin{equation}\label{eqn:toSUGRAfields}
  \begin{aligned}
    \begin{pmatrix} g_{ij} - B_{ik} g^{kl} B_{lk} & -B_{ik} g^{kj} \\
      g^{ik} B_{kj} & g^{ij}
    \end{pmatrix} &= 
    \widehat{\mathcal{H}}^{\hat I\hat J} = \widehat{E}_A{}^{\hat I} \mathcal{H}^{AB} \widehat{E}_A{}^{\hat J} \\
    \phi &= d + \frac14 \log | \det g | + \frac12 \log \det | \tilde e |
  \end{aligned}
\end{equation}
where $g_{ij}$, $B_{ij}$ and $\phi$ are the metric, $B$-field and dilaton on the physical target space $M$. To obtain $\phi$ from the generalized dilaton $d$, we also need $\tilde e$ which is a $D\times D$ matrix with the components $\tilde e_{a\tilde i} = \llangle T_a , \tilde h^{-1} \partial_{\tilde i} \tilde h\rrangle$. After plugging these relations into the action \eqref{eqn:Sdftwzw} and taking $\tilde H$ to be unimodular, one obtains the NS/NS action
\begin{equation}\label{eqn:NSSUGRA}
  S_\mathrm{NS} = \int\mathrm{d}^{D}x\,
    \sqrt{|\det g|} e^{-2\phi} \big(\mathrm{Ric} + 4 \partial_i \phi \partial^i \phi 
    - \frac{1}{12} H_{ijk} H^{ijk} \big)
\end{equation}
for type II SUGRA with the curvature scalar $\mathrm{Ric}$ and the three-form flux $H_{ijk} = 3 \partial_{[i} B_{jk]}$.

If $\tilde H$ is not unimodular this identification fails due to subtleties with integration by parts. However, one can apply the same procedure directly at the level of the field equations. They arise by varying $S_\mathrm{NS}$ with respect to the generalized dilaton and the generalized metric. We discuss them in more detail in section~\ref{sec:applications}. Here, we just need that they eventually give rise to
\begin{equation}\label{eqn:genSUGRAeom}
\begin{aligned}
  \mathrm{Ric}_{mn}   -\frac{1}{4}H_{mpq}H_{n}{}^{pq} + \nabla_m X_n + \nabla_n X_m &= 0 \\
  \nabla^p H_{pmn} - X^p H_{pmn} - \partial_m X_n + \partial_n X_m  &=0 \\
  \mathrm{Ric}  + 4 \nabla_n X^n -  4 X_n X^n - \frac{1}{12} H_{mnp} H^{mnp} &=0
\end{aligned}
\end{equation}
after one substitutes the generalized metric and generalized dilaton with \eqref{eqn:toSUGRAfields}. $\mathrm{Ric}_{mn}$ is the Ricci tensor and the form $X_m$ is defined by
\begin{equation}
  X_m = \partial_m \phi + I^n ( g_{nm} + B_{nm} ) - V_m \,.
\end{equation}
These equations are imposed by $\kappa$-symmetry on the worldsheet and go under the name generalized SUGRA \cite{Arutyunov:2015mqj}. For vanishing $I^m$ and $V_m$, we also have conformal symmetry and \eqref{eqn:genSUGRAeom} are field equations for \eqref{eqn:NSSUGRA}. This is only the case if $\tilde H$ is unimodular. In general we find
\begin{equation}
  \widehat{X}_{\hat M} = \begin{pmatrix} I^m \\ - V_m \end{pmatrix} = \frac12 \left( 
  \partial_{\hat I} \log|\det e | + \partial_{\hat I} \widehat{E}_A{}^{\hat I} \widehat{E}^A{}_{\hat M} \right) \quad \text{with} \quad T_a e^a{}_i = m^{-1} \partial_i m \,.
\end{equation}
There is a lot more to say about generalized SUGRA, its relation to DFT \cite{Sakatani:2016fvh,Sakamoto:2017wor} and the role it plays in string theory. Here we just want to highlight that it is automatically built into \DFTwzw{}.

We have now established that \DFTwzw{} is indeed the low energy effective target space theory of the $\mathcal{E}$-model. This is interesting from a conceptual point of view, but it also allows us to make an interesting conjecture. Remember that identifying the action \eqref{eqn:Sdftwzw} with (generalized) supergravity is based on a generalized frame field. Thus one should assume that there exists a generalized frame field on $M=\tilde H \backslash \mathdsl{D}$ for every choice of $\mathdsl{D}$ and $\tilde H$ which are compatible with the $\mathcal{E}$-model. Indeed this is the case. Even better there is an explicit construction for $\widehat{E}_A{}^{\hat I}$. It is extremely useful in the context of consistent truncations with half-maximal supersymmetry where this construction was an open problem for some time. We do not present the rather technical detail here and instead refer to \cite{Demulder:2018lmj}. An explicit example is presented in section~\ref{sec:applications}.

\section{The Ramond/Ramond sector}
In addition to the NS/NS sector in the previous subsection, the bosonic field content of type II string theory includes a Ramond/Ramond (R/R) sector. Almost all physically relevant target spaces carry non-trivial R/R fluxes. Thus if one wants to apply Poisson-Lie T-duality to them, for example as a solution generating technique, this sector plays a very important role. Studying it directly on the worldsheet is involved. A popular approach to $\lambda$- and $\eta$-deformations is the Green-Schwarz string \cite{Borsato:2016ose}. But to the best of our knowledge there is no full $\mathcal{E}$-model description containing R/R fields yet. Fortunately, the situation in DFT is better \cite{Hohm:2011dv,Jeon:2012kd}. There are even two different approaches for an abelian T-duality covariant description of massless R/R potentials. Here we focus on the one \cite{Hohm:2011dv} which treats them as an O($D$,$D$) Majorana-Weyl (MW) spinor.

To make contact with the $\mathcal{E}$-model in the last section, we rely on insights from CSFT calculations based on a WZW-model. One could in principal try to follow a similar route here but technically that would be very involved. We rather prefer a much simpler approach based on symmetry arguments to derive the R/R sector action of \DFTwzw{} and its gauge transformations.

\subsection{Action and symmetries}
A key ingredient in this construction are O($D$,$D$) MW spinors on the Lie group $\mathdsl{D}$. They come with $\Gamma$-matrices which are governed by the Clifford algebra
\begin{equation}
  \{ \Gamma_A, \Gamma_B \} = 2 \eta_{AB}
\end{equation}
and admit charge conjugation defined by
\begin{equation}
  C \Gamma_A C^{-1} = (\Gamma_A)^\dagger\,.
\end{equation}
If we decompose the anti-commutator into the three contributions $\{\Gamma_a, \Gamma_b\}$=0, $\{\Gamma^a, \Gamma^b\}$=$0$ and $\{\Gamma^a, \Gamma_b\}$=$2 \delta^a_b$, we recognize that $\Gamma^a$ and $\Gamma_a$ are nothing else than fermionic creation and annihilation operators. They are chosen such that $(\Gamma_a)^\dagger$=$\Gamma^a$ holds. If we further introduce a vacuum $|0\rangle$ that is annihilated by all $\Gamma_a$, each O($D$,$D$) MW spinor admits the expansion
\begin{equation}\label{eqn:formstospinor}
  \chi = \sum\limits_{p=0}^D \frac1{2^{p/2} \, p!} C^{(p)}_{a_1\dots a_p} \Gamma^{a_1} \dots \Gamma^{a_p} | 0 \rangle
\end{equation}
into differential forms. The Weyl in MW tells us that the spinor is either chiral or anti-chiral. Equivalently, we only have even degree forms (chiral) or odd degree forms (anti-chiral). We will see later that these forms exactly capture the R/R gauge potentials of type IIB/IIA. 

We already encounter O($D$,$D$) transformations $\mathcal{O}^A{}_B$ acting on a vector $V^A \rightarrow \mathcal{O}^A{}_B V^B$ or on more general tensors. But how do they act on the spinor $\chi$? It transforms as $\chi \rightarrow S_\mathcal{O} \chi$ where $S_\mathcal{O}$ is implicitly defined by
\begin{equation}
  S_\mathcal{O} \Gamma^A S_\mathcal{O}^{-1} = \mathcal{O}^A{}_B \Gamma^B \,.
\end{equation}
For infinitesimal transformation $\delta V^A = \delta \mathcal{O}^A{}_B V^B$  this prescription gives rise to
\begin{equation}
  \delta \chi = \frac14 \delta \mathcal{O}_{AB} \Gamma^{AB} \chi
\end{equation}
with the notation $\Gamma^{AB} = \Gamma^{[A} \Gamma^{B]}$.  From this relation we derive the spinor action of the $\eta$-metric compatible covariant derivatives
\begin{equation}\label{eqn:nablatspinor}
  \nabla^{(t)}_A \chi = \mathdsl{E}_A{}^I \partial_I \chi - \frac{t}4 \mathdsl{F}_{ABC} \Gamma^{BC} \chi - w(\chi) \mathdsl{E}_A{}^I \partial_I \log | \det \mathdsl{E} | \chi\,.
\end{equation}
The $\Gamma$-matrices are chosen such that they are covariant constant. That means like $\eta_{AB}$ they are annihilated by $\nabla^{(t)}_A$.

Remember that we have distinguished one member of the family \eqref{eqn:nablatspinor}, $\nabla_A = \nabla_A^{(1/3)}$, to write down the \DFTwzw{} action for the NS/NS sector \eqref{eqn:Sdftwzw} in $2D$-diffeomorphisms manifest way. It looks very similar to the HHZ action \cite{Hohm:2010pp} but with all partial derivatives replaced by covariant ones. Let us postulate that the same also applies to the R/R sector action. In this case we look at the action in \cite{Hohm:2011dv} and obtain
\begin{equation}\label{eqn:SdftwzwR}
  S_\mathrm{R} = \frac14 \int \dd^{2D} X \overline{\slashed{\nabla}\chi} \mathcal{K} \slashed{\nabla} \chi \quad \text{with} \quad \mathcal{K} = C^{-1} S_\mathcal{H}
\end{equation}
with the spinor conjugation $\overline{\chi} = \chi^\dagger C$. For an invariant action the density of the integrand has to be one. Thus, $w(\chi)$ has to be 1/2 (because $\chi$ appears two times under the integral). After fixing this last freedom, one finds that this action has all the symmetries of its NS/NS counter part. More explicitly, there are
\begin{itemize}
  \item {\it Global O($D$,$D$) transformations} mediated by
    \begin{equation}
      \xi \rightarrow S_\mathcal{O} \xi\,.
    \end{equation}
  \item {\it Conventional 2D-diffeomorphisms} mediated by the Lie derivative
    \begin{equation}
      \delta \chi = L_\xi \chi = \xi^B D_B V^A + \frac12 D_B \xi^B \chi\,.
    \end{equation}
  \item {\it Generalized diffeomorphisms} mediated by the generalized Lie derivative
    \begin{equation}
      \delta \chi = \mathcal{L}_\xi = \xi^A \nabla_A \chi + \frac12 \nabla_A \xi_B \Gamma^{AB} \chi + \frac12 \nabla_A \xi^A \chi\,.
    \end{equation}
\end{itemize}
Again the first two close without any restriction and are manifest, while generalized diffeomorphisms require the section condition \eqref{eqn:SC} to close and leave the action invariant.

From the expansion, we see that we have all even/odd forms from zero to $D$. This is the same situation as in the democratic formulation of type II SUGRA. There one has to cut down the number of degrees of freedom by imposing a duality condition involving the Hodge star. Here the same happens if we just leave the MW spinor $\chi$ unconstrained, we would end up with too many dynamical fields. Hence, we additionally impose the constraint
\begin{equation}\label{eqn:dualitycond}
  G  = - \mathcal{K} G \quad \text{with} \quad G = \slashed{\nabla} \chi\,.
\end{equation}
$G$ combines all R/R field strengths in one O($D$,$D$) MW spinor. By varying the action \eqref{eqn:SdftwzwR} with respect to $\chi$, we find its field equation
\begin{equation}\label{eqn:eomG}
  \slashed{\nabla} ( \mathcal{K} G ) = 0 \,.
\end{equation}
They are completed with the Bianchi identity
\begin{equation}\label{eqn:BianchiG}
  \slashed{\nabla} G = 0
\end{equation}
which follows for the fact that $\chi$ has to satisfy the section condition. In this case one can easily check that $\slashed{\nabla}^2 \chi$=0 holds.

\subsection{Equivalence to (generalized) SUGRA}
In contrast to the NS/NS sector action we have not derived \eqref{eqn:SdftwzwR} from CSFT theory. Rather we have made an educated guess based on symmetries and results from conventional DFT. To check that we did not do anything wrong, we now show that the field equations we derived from this action match the ones from generalized SUGRA. There are two good reasons to consider the field equations instead of the action. First, they are very simple and second for the most general case where $\tilde H$ is not unimodular there is no action on the physical target space $M=\tilde H \backslash \mathdsl{D}$. We encountered this problem already for the NS/NS sector and again it originates from subtleties with integration by parts.

We use the ansatz~\eqref{eqn:toSUGRAfields} for the generalized metric and the corresponding version for a MW spinor of weight $w(G)=1/2$,
\begin{equation}
  \sum\limits_{p=0}^D \frac1{2^{p/2} \, p!} G^{(p)}_{i_1\dots i_p} \widehat{\Gamma}^{i_1} \dots \widehat{\Gamma}^{i_p} | 0 \rangle = \widehat{G} = \frac1{\sqrt{|\det \tilde e |}}
  S_{\widehat{E}} G \,.
\end{equation}
As before the generalized frame field $\widehat{E}_A{}^{\hat I}$ is essential in this identification. It is an O($D$,$D$) element and therefore leaves the $\Gamma$-matrices invariant. So even if we write them here with a hat to be consistent with the notation, as tensors they are equivalent to the unhatted version in flat indices. The same happens for $\eta_{AB}$ and $\widehat{\eta}_{\hat I\hat J}$ in figure~\ref{fig:indices}. To avoid coping with every single differential form, it is convenient to combine them into the polyform
\begin{equation}
  F = e^{\phi - B} \sum\limits_{p=0}^D G^{(p)}\,.
\end{equation}
Here $B$ denotes the two-form $B$-field and $\phi$ is the dilaton. Under this identification the field equations \eqref{eqn:eomG} and the Bianchi identity \eqref{eqn:BianchiG} become
\begin{equation}
  \mathbf{d} ( \star F ) = 0 \qquad \mathbf{d} F = 0
\end{equation}
with the twisted exterior derivative
\begin{equation}
  \mathbf{d} = \dd + (H - \dd \phi + \iota_I B - V )\wedge + \iota_I\,.
\end{equation}
But this is nothing else than a very compact form of the R/R sector generalized SUGRA field equations \cite{Arutyunov:2015mqj}. The first one is automatically solved after imposing the duality condition \eqref{eqn:dualitycond}.

This result is very nice because it shows that the connection between DFT and the $\mathcal{E}$-model is mutually beneficial. In the last section our approach was more guided by the $\mathcal{E}$-model perspective. It led to a construction of generalized frame fields which is very valuable for DFT and generalized geometry. Here we went into the opposite direction. The derivation in the R/R sector exploits the DFT construction \cite{Hohm:2011dv} to learn more about the $\mathcal{E}$-model. We follow this idea further in the next section to eventually obtain the transformation rules for R/R field under full Poisson-Lie T-duality.

\section{Poisson-Lie symmetry and T-duality in target space}\label{sec:PLsymts}
There is still one small puzzle left in the connection between \DFTwzw{} and the $\mathcal{E}$-model. It is related to the generalization we have made in the last bullet point on page~\ref{bullet:genEmodel}. This point implies that there is something special about constant $\mathcal{E}$'s. Indeed, we would not be able to bring the time evolution of the current $j(\sigma)$ into the form \eqref{eqn:ddtj} if $\mathcal{E}$ depends on the worldsheet coordinates. But why should we care? Because \eqref{eqn:ddtj} is equivalent to having a $\mathfrak{d}$-valued flat connection
\begin{equation}\label{eqn:J}
  J(\sigma) = j_+(\sigma) \dd \xi^+ + j_-(\sigma) \dd \xi^- \qquad j_\pm(\sigma) = \frac12 ( \mathcal{E} \pm  1 ) j(\sigma) \,.
\end{equation}
It is not hard to check that \eqref{eqn:ddtj} is equivalent to
\begin{equation}\label{eqn:flatcurrent}
  \dd J(\sigma) - [J(\sigma), J(\sigma)] = 0\,.
\end{equation}
This is exactly the flat connection we encountered in section~\ref{sec:Emodel} as can be readily verified by identifying the space- and time-components of $J(\sigma)$ with respectively $j(\sigma)$ and $ \mathcal E j(\sigma)$. The current $J(\sigma)$ thus contains the ingredients for the Lax connection \eqref{eqn:laxconnection} that ensures the integrability of the $\mathcal{E}$-model. Note that this time however $J(\sigma)$, in contrast to the current $j(\sigma)$, has $2D$ independent components instead of just $D$. It also does not incorporate the spectral parameter $\lambda$. Thus, integrability of the $\sigma$-model does not automatically follow from \eqref{eqn:flatcurrent}.

Let us look at a simple example to get a better idea of $J$'s physical interpretation and how the existence of a current \eqref{eqn:flatcurrent} relates to T-duality. For a closed string on a circle we have a two-dimensional, abelian Lie algebra $\mathfrak{d}$ spanned by the two commuting generators $T$ and $\tilde T$. Each generates a maximally isotropic subgroup of $\mathdsl{D}$ whose group elements are
\begin{equation}
  \mathdsl{g} = e^{i \tilde x \tilde T} e^{i x T} \,.
\end{equation}
For a constant $\mathcal{E}$ we figured out that \eqref{eqn:J} gives rise to a flat connection $J$. Thus, the integral over $J$
\begin{equation}
  \mathdsl{D} \ni Q = P \exp \int_\gamma J
\end{equation}
is invariant under continues deformations of the path $\gamma$. It is particularly simple to evaluate if we choose $\gamma$ to be at a fixed time $\tau$ and $\sigma\in[0,2\pi]$
\begin{equation}
  Q = P \exp \int\limits_0^{2\pi} \dd \sigma\, j(\sigma) =  e^{i \left( \tilde x(\tau, 2\pi) - \tilde x(\tau, 0) \right) \tilde T}
      e^{i \left( x(\tau, 2\pi) - x(\tau, 0) \right) T} \,.
\end{equation}
A conserved group element $Q$ implies that the two quantities
\begin{equation}
  x(\tau, 2\pi) - x(\tau, 0) = 2 \pi w  \qquad \text{and} \qquad
  \tilde x(\tau, 2\pi) - \tilde x(\tau, 0) = 2 \pi m
\end{equation}
are conserved independently. $w$ and $m$ are nothing else than the winding and momentum numbers of the closed string. Because it is closed, both have to be integer valued. T-duality just exchanges $x\leftrightarrow\tilde x$ and winding$\leftrightarrow$momentum. Intriguingly this argument does not require $\mathdsl{D}$ to be abelian. It applies to all $\mathcal{E}$-models \cite{Klimcik:1996nq}. However, one needs to know the global properties of $\mathdsl{D}$. For abelian T-duality this information is available. But already for non-abelian T-duality it is much more subtle as we explain in one of the open problems presented in section~\ref{sec:conlusions}.

From Noether's theorem we know that each conserved current generates a symmetry. For $J$ this symmetry was dubbed Poisson-Lie symmetry \cite{Klimcik:1995jn}. The name originates from the fact that a large class target spaces with this symmetry are Poisson-Lie groups. It is the necessary condition for a target space to admit Poisson-Lie T-duality but it is not sufficient. Additionally the associated Lie group $\mathdsl{D}$ must have more than one maximally isotropic subgroups. As our small example shows, abelian T-duality is a special case where $\mathdsl{D}$ is abelian.

As Poisson-Lie symmetry is essential to the $\mathcal{E}$-model, it should also be important in \DFTwzw{}. The key to understand what is going on is to look at two different solutions of the section condition for the same doubled Lie group $\mathdsl{D}$. As shown in section~\ref{sec:solSC} such solutions arise from different maximally isotropic subgroups, let us call them $\tilde H$ and $\tilde H'$. For both, we take a look at the three important ingredients in the construction
\begin{center}
  \setlength{\tabcolsep}{2em} 
  \begin{tabular}{ccc}
    physical target space & group element & generators \\[0.5em]
    $\tikzmarkMath{TS1}{m \in \tilde H \backslash \mathdsl{D}}$ &
    $\tikzmarkMath{g}{\mathdsl{g}(X^I) = \tilde h(\tilde x^{\tilde i}) m(x^i)}$ &
    $\tikzmarkMath{T}{\mathdsl{T}_A = \begin{pmatrix} \tilde T^a & T_a \end{pmatrix}}$ \\[3em]
    $\tikzmarkMath{TS2}{m' \in\tilde H' \backslash \mathdsl{D}}$ &
    $\tikzmarkMath{gp}{\mathdsl{g}(X'^I) = \tilde h'(\tilde x'^{\tilde i}) m'(x'^i)}$ &
    $\tikzmarkMath{Tp}{\mathdsl{T}'_A = \begin{pmatrix} T'_a & \tilde T'^a \end{pmatrix}}$\,.
  \end{tabular}%
  \tikz[overlay,remember picture]{
    \draw[<->] (TS1) -- (TS2) node[midway,anchor=west] {Poisson-Lie T-d.};
    \draw[<->] (g) -- (gp) node[midway,anchor=west] {conventional $2D$-diffs};
    \draw[<->] (T) -- (Tp) node[midway,anchor=west] {global O($D$,$D$)};
  }%
\end{center}
The $\mathcal{E}$-model tells us that the group elements $\mathdsl{g}$ actually do not change. Only their parameterization does. But this coordinate change is just a conventional, volume preserving $2D$ diffeomorphism. A similar argument holds for the generators which are rotated by a global O($D$,$D$) transformation such that $\eta_{AB}$ is preserved. To keep the action of \DFTwzw{} invariant under these transformations, the generalized metric $\mathcal{H}_{AB}$, dilaton $d$ and the MW spinor $\chi$ have to be transformed accordingly. How this works is discussed in full detail in the previous sections. But there is an important subtlety we have to take care of. We start with the physical target space $\tilde H \backslash \mathdsl{D}$. Due to the section condition $\mathcal{H}_{AB}$ is only allowed to depend on the coordinates $x^i$. Next, a conventional $2D$ diffeomorphism is performed to express all fields in terms of the new coordinates $X'^I$. But without any further restriction the generalized metric would not just depend on the $x'^i$'s. In general it also picks up a dependence on the section condition violating $\tilde x'^{\tilde i}$'s. This is not surprising because we know that a generalized metric without any further restrictions can capture all possible target space geometries and thus has to violate Poisson-Lie symmetry. In general this rules out Poisson-Lie T-duality. A notable exception is when $\mathcal{H}_{AB}$ is constant. At the same time $\mathcal{E}$ is of course constant too. We now understand why Poisson-Lie T-duality is not manifest in the generalized metric formulation of Hohm, Hull and Zwiebach \cite{Hull:2009mi,Hohm:2010pp}. It lacks conventional 2$D$-diffeomorphisms. This is not a problem for totally abelian Lie groups $\mathdsl{D}$, because in this case they only act like the global O($D$,$D$) transformations. But if $\mathdsl{D}$ becomes non-abelian, which always happens if we are interested in anything else than abelian T-duality, this additional local symmetry becomes essential.

\begin{figure}[!t]
  \centering
  \begin{tikzpicture}
    \draw[fill=black!20] (0.5,0) ellipse (3cm and 2cm) node[xshift=-2.5cm] {$\mathdsl{D}$};
    \draw[fill=white] (0.5,0) ellipse (2cm and 1.5cm) node[yshift=1.2cm] {$\mathdsl{I}$};
    \draw (0,0) circle (1cm) node[xshift=-0.6cm] {$\tilde H$};
    \draw (1,0) circle (1cm) node[xshift=0.6cm] {$\tilde H'$};
    \draw (3,0) -- (5,1) node[anchor=west] {spectators};
  \end{tikzpicture}
  \caption{If the group $\mathdsl{I}$ of isometries which act from the right on $\mathdsl{D}$ has at least two different, maximally isotropic subgroups (here $\tilde H$ and $\tilde H'$) we encounter Poisson-Lie T-duality with spectators.}
  \label{fig:PLspectators}
\end{figure}
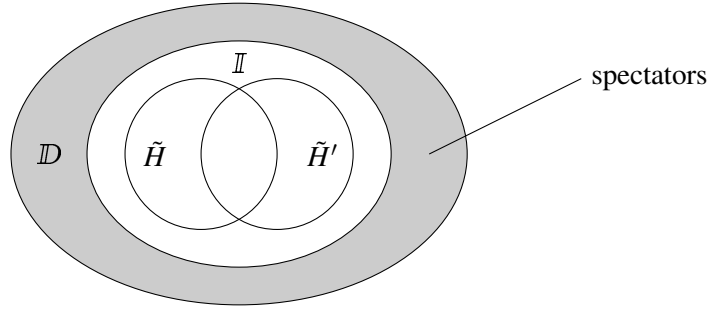

There are two advantages of the target space perspective. Most important, the discussion for the generalized metric applies for the dilaton and the R/R fields, too. But in contrast to $\mathcal{H}_{AB}$ they are much harder to access on the worldsheet. We come back to this point at the end of this section. Furthermore, the condition for Poisson-Lie symmetry has a nice geometric interpretation as isometries for the field on $\mathdsl{D}$. This becomes obvious if we write the condition that $\mathcal{H}_{AB}$ has to be constant in terms of the conventional Lie derivative
\begin{equation}\label{eqn:isometryHAB}
  L_{\mathdsl{E}_A} \mathcal{H}_{BC} = D_A \mathcal{H}_{BC} = 0
\end{equation}
where $\mathdsl{E}_A$ are the right invariant vector fields on $\mathdsl{D}$. Hence, this equation tells us that Poisson-Lie symmetry requires a freely acting isometry group $\mathdsl{D}$ on the doubled space. Isometries are always closely connected to T-duality. Thus, Poisson-Lie T-duality feels at the first glace a bit strange because it works for certain target spaces which does not seem to have any isometries. Here the doubled formulation is much more democratic. It is interesting to note that there exists a slight generalized of the condition \eqref{eqn:isometryHAB}. If we start from a Lie group $\mathdsl{D}$ with a maximally isotropic subgroup $\tilde H$, the group $\mathdsl{I}$ of isometries which act from the right on $\mathdsl{g}\in\mathdsl{D}$ has to be at least $\tilde H$. Otherwise we would not be able to solve the section condition. On the other hand, $\mathdsl{I}$=$\mathdsl{D}$ is the situation with the most symmetry captured by \eqref{eqn:isometryHAB}. In general we expect to find
\begin{equation}
  \tilde H \subseteq \mathdsl{I} \subseteq \mathdsl{D}\,.
\end{equation}
If there is at least one other maximally isotropic subgroup in $\mathdsl{I}$, we are able to apply Poisson-Lie T-duality. But this time the part $\mathdsl{I}\backslash\mathdsl{D}$ is just a spectator and does not change. This situation is depicted in figure~\ref{fig:PLspectators}.

Looking at the analog version of \eqref{eqn:isometryHAB} for the generalized dilaton, 
\begin{equation}
  L_{\mathdsl{E}_A} e^{-2 d} = 0\,,
\end{equation}
constraints a Poisson-Lie symmetric dilaton to be of the form
\begin{equation}\label{eqn:conddilaton}
  \phi = \frac14 \log | \det g | - \frac12 \log | \det e | +  \phi_0 \,,
\end{equation}
with $\phi_0$ a constant. Under Poisson-Lie T-duality $\phi_0$ is invariant. Hence, we immediately read off the transformation rule
\begin{equation}
  \phi' = \phi + \frac14 \log \left| \frac{\det g'}{\det g} \right| + \frac12 \log \left| \frac{\det e}{\det e'} \right|
\end{equation}
for the dilaton in two target spaces connected by Poisson-Lie T-duality. It matches with the recent result \cite{Jurco:2017gii} obtained from the contraction of Courant algebroids. For the R/R fields the same procedure applies. First we impose the isometry constraint
\begin{equation}
  L_{\mathdsl{E}_A} G = 0
\end{equation}
which simply implies that
\begin{equation}\label{eqn:condG0}
  \widehat{G} = \sqrt{| \det e |} S_{\widehat{E}} G_0
\end{equation}
where $G_0$ is a constant O($D$,$D$) MW spinor. Again, $G_0$ is invariant under Poisson-Lie T-duality and therefore 
\begin{equation}\label{eqn:trafoRR}
  \widehat{G}' = \sqrt{\left| \frac{\det e'}{\det e} \right|} S_{\widehat{E}' \widehat{E}^{-1}} \widehat{G}
\end{equation}
captures the transformation of R/R fluxes. Non-abelian T-duality is a special case where $\mathdsl{D}$ is just $T^* G$=$G \ltimes \mathbb{R}^D$ of a $D$-dimensional group $G$. In this case, the R/R transformation rule were already known \cite{Sfetsos:2010uq}. Finding them was an important step to establish non-abelian T-duality as a solution generating technique and has trigger a lot of activity. However an analog version for the full Poisson-Lie T-duality was missing. \DFTwzw{} allows to derive it explicitly. Besides the explicit construction of generalized frame fields, this is the second important outcome of the target space approach to Poisson-Lie T-duality we present in this paper.

\section{Applications}\label{sec:applications}
We have already seen two important achievements of a target space approach to the $\mathcal{E}$-model. First we are able to find a construction for generalized frame fields which are essential for generalized Scherk-Schwarz reductions. Second, we are able to derive the transformation rules for the R/R sector fields. This is particularly important to apply Poisson-Lie T-duality as a solution generating technique. Both are more on the conceptual side. But there is also a big advantage for finding solutions of the field equations of (generalized) SUGRA: for backgrounds with Poisson-Lie symmetry, they simplify drastically and become purely algebraic.

To derive these algebraic field equations, we have to vary the action with respect to the generalized metric and generalized dilaton. For the former there is a small subtlety, because $\mathcal{H}_{AB}$ is not just a symmetric rank two tensor like in general relativity but further restricted to be an O($D$,$D$) element. This requires to introduce additional projectors \cite{Hohm:2010pp}. We will not worry about the details here, but rather present the final result. From the variation with respect to the generalized dilaton, we obtain
\begin{equation}\label{eqn:genricci}
  \mathcal{R} = \frac1{12} \mathdsl F_{ACE} \mathdsl F_{BDF} \left( 3 \mathcal{H}^{AB} \eta^{CD} \eta^{EF} - \mathcal{H}^{AB} \mathcal{H}^{CD} \mathcal{H}^{EF} \right) = 0
\end{equation}
for a Poisson-Lie symmetric background. $\mathcal{R}$ is the generalized Ricci scalar. It is closely related to the scalar potential which arises in gauged SUGRAs obtained from a generalized Scherk-Schwarz reduction. Most important, there are no derivatives in \eqref{eqn:genricci}. It is just a cubic equation in $\mathcal{H}_{AB}$. Variation with respect to the generalized metric (and applying the appropriate projectors) gives rise to
\begin{equation}\label{eqn:eomHAB}
  \mathcal{R}_{AB} - \frac{e^{2\phi_0}}{16} \mathcal{H}_{(A}{}^C G_0^\dagger \Gamma_{B)C} \mathcal{K} G_0   = 0
\end{equation}
with the generalized Ricci tensor
\begin{equation}
  \mathcal{R}_{AB} = \frac18 (\mathcal{H}_{AC} \mathcal{H}_{BF} - \eta_{AC} \eta_{BF} ) (\mathcal{H}^{KD} \mathcal{H}^{HE} - \eta^{KD} \eta^{HE})\mathdsl F_{KH}{}^C\mathdsl F_{DE}{}^F \,.
\end{equation}
A solution to the SUGRA fields equations guarantees that the one-loop $\beta$-function on the worldsheet vanishes. Hence, it is not surprising the that $\mathcal{R}_{AB}$ also appears in the RG flow calculation for a double sigma model presented in \cite{Avramis:2009xi,Sfetsos:2009vt}. The quantities which are governed by the field equations are $\mathcal{H}_{AB}$, $\phi_0$ and $G_0$. Remember that these are exactly the ones which are invariant under Poisson-Lie T-duality. Thus, at this level it is completely manifest that Poisson-Lie T-duality is indeed a solution generating technique. Establishing this result at the level of a target space theory without manifest Poisson-Lie symmetry, like SUGRA or Hohm, Hull and Zwiebach DFT, is much more involved \cite{Sakatani:2019jgu,Catal-Ozer:2019hxw}.

To provide an explicit example for the ideas presented in this paper, we come back to the integrable $\eta$-deformation from section~\ref{sec:Emodel}.  We already know its $\mathcal{E}$ operator and thus it is not hard to obtain the generalized metric \cite{Klimcik:2015gba}
\begin{equation}\label{eqn:genmetricetadef}
  \mathcal{H}^{AB} = \begin{pmatrix} \kappa k_{ab} & - \kappa k_{ac} R^{cb} \\
    \kappa R^{ac} k_{cb} & \frac{\displaystyle k^{ab}}{%
    \displaystyle\kappa}  - \kappa R^{ac} k_{cd} R^{db}
  \end{pmatrix}\,,
\end{equation}
in the basis \eqref{eqn:basisdeta} of the Lie algebra $\mathfrak{d}$. If we want to make contact with the metric and the $B$-field by applying \eqref{eqn:toSUGRAfields}, we further need the generalized frame field
\begin{equation}\label{eq:GFFeta}
  \widehat{E}_A{}^{\hat I }= \begin{pmatrix} 
    e^a{}_i & \Pi^{ab} e_b{}^i\\
    0       &  e_a{}^i
  \end{pmatrix} \,.
\end{equation}
We already encountered $e^a{}_i$ a few times before. It is defined by the left-invariant Maurer Cartan form $T_a e^a{}_i = m^{-1} \partial_i m$ and also gives rise the right invariant vector fields $e_a{}^i$ (the inverse transposed of $e^a{}_i$). Finally, there is the Poisson-Lie structure $\Pi^{ab}$ for the Lie group $G$=$\tilde{H}\backslash\mathdsl{D}$. It can be derived from the $R$-matrix
\begin{equation}
  \Pi^{ab} = R^{ab} - R^{cd} M_c{}^a M_d{}^b
    \qquad \text{with} \qquad
  M_a{}^b T_b = m T_a m^{-1} \,.
\end{equation}

If one checks the field equation for the generalized dilaton \eqref{eqn:genricci} one finds that it is impossible to solve it for a simple group $G$ as physical target space. Instead one has to combine at least a compact and a non-compact simple factor. The simplest setup along those lines is G=SL(2)$\times$SU(2) with the Killing form $k_{ab}=\mathrm{diag}(-1,+1,+1,+1,+1,+1)$. It gives rise to the $\eta$-deformation of AdS$_3\times$S$^3$. A ten dimensional background arises when we add an additional $T^4$ factor. If we now check the field equations for the generalized metric \eqref{eqn:eomHAB} we see that they are violated without any R/R fluxes $G_0$. Thus, the question arises if it is possible to switch on a Poisson-Lie symmetric $G_0$ such that we find a solution to generalized SUGRA. We have to work in the framework of generalized SUGRA because
\begin{equation}
  I^i = \frac12 ( R^{bc} - \Pi^{bc} ) f_{bc}{}^a e_a{}^i
\end{equation}
does not vanish (while $V_i$=$0$). This question can be answered in the affirmative. We will not go into the technical details here. They are presented in \cite{Demulder:2018lmj}. Having $\mathcal{H}^{AB}$, $\phi_0$, $G_0$ and the generalized frame field $\widehat{E}_A{}^{\hat I}$, one obtains the SUGRA fields from \eqref{eqn:toSUGRAfields}, \eqref{eqn:conddilaton} and \eqref{eqn:condG0}.

We on purpose skip this last step here because even if the calculation are mechanical and can be done by computer algebra software they are very cumbersome. A big advantage of \DFTwzw{} is that we can discuss all important properties of a Poisson-Lie symmetric background avoiding this step.

\section{Conclusions}\label{sec:conlusions}
We have reviewed Poisson-Lie symmetry and T-duality \cite{Klimcik:1995ux,Klimcik:1995dy,Klimcik:1995jn}. On the worldsheet they are most clearly captured in the language of the $\mathcal{E}$-model \cite{Klimcik:1995dy,Klimcik:1996nq,Klimcik:2015gba}. One of their important applications is the construction of new integrable two-dimensional $\sigma$-models. Looking at the Yang-Baxter $\sigma$-model \cite{Klimcik:2002zj,Klimcik:2008eq}, we explored the ingredients and ideas behind this approach. The insights we gained from this tractable example are quite general and also apply to more complicated multiparameter deformations.

Once we start to appreciate the $\mathcal{E}$-model, the natural question that arises is: What is the associated low energy effective theory in target space? We discussed that the obvious answer, generalized SUGRA \cite{Arutyunov:2015mqj}, is correct. But it is not completely satisfying because it obfuscates the powerful structure underlying Poisson-Lie symmetry which we exploited to construct integral deformations. Thus, we instead established the relation to \DFTwzw{} \cite{Blumenhagen:2014gva,Blumenhagen:2015zma}. It was originally derived to better understand the significance of a DFT description for target spaces which do not admit closed string winding modes. A canonical example is the three sphere with $H$-flux or on the worldsheet the SU(2) WZW-model. Here every string can be contracted to a point and thus there are no winding modes. Treating left- and right-moving sector independently, one is still able to derive a double field theory on these backgrounds, which was for obvious reasons dubbed \DFTwzw{}. This theory is formulated on a Lie group whose dimension is two times the number of the target space dimension and has some additional structure compared to the standard generalized metric formulation due to Hohm, Hull and Zwiebach \cite{Hull:2009mi,Hohm:2010pp}. Exactly this additional structure allows for a perfect match with the $\mathcal{E}$-model as we show in section~\ref{sec:DFTwzwNSNS}. From our reasoning in the beginning we know that this theory has to be equivalent to generalized SUGRA. Indeed we are able to prove this conjecture with the help of a generalized frame fields. The latter is essential for generalized Scherk-Schwarz reductions in DFT \cite{Grana:2012rr,Geissbuhler:2011mx,Aldazabal:2011nj,Catal-Ozer:2017cqr} and in the construction of consistent truncations to (half) maximal gauged SUGRAs. That makes generalized frame fields powerful objects. Unfortunately an explicit construction for them was lacking. Instead they had to be guessed such that they satisfy three defining relations. One of them is a partial differential equation which makes this business quite hard. Luckily with the insights from \DFTwzw{} and the $\mathcal{E}$-model it is possible to provide an explicit construction.

Up to this point, the results benefit mostly our understanding of doubled target space theories. We now know that winding modes are not strictly required to make sense of a DFT. The $\mathcal{E}$-model rather suggests that the doubled space should be interpreted as a phase space which appears naturally in the first order formulation on the worldsheet. Furthermore we are now able to construct all O($D$,$D$) generalized frame fields explicitly. But there is an important lesson to be learned from DFT too. Even if it is not a symmetry of full closed string theory (to all orders in $\alpha'$ and $g_s$), non-abelian T-duality is very useful as a solution generating technique. A key to this development was to understand how R/R fluxes transform under it \cite{Sfetsos:2010uq}. If we want to do the same with full Poisson-Lie T-duality, which includes non-abelian T-duality as a special case, we have to address the same question. At this point insights from DFT \cite{Hohm:2011dv} are very helpful. Adapting the treatment of R/R fluxes as O($D$,$D$) MW spinor to \DFTwzw{} allowed us to derive these transformations. It shows that the effort to make T-duality manifest at the level of the target space is not just a purely academic exercise. Instead it allows one to really derive new results which were not accessible before. At the same time it becomes clear that Poisson-Lie T-duality is indeed a solution generating technique for generalized SUGRA if we look at the field equations \eqref{eqn:genricci} and \eqref{eqn:eomHAB}. They are equivalent to the ones in generalized SUGRA but written in terms of the quantities $\mathcal{H}_{AB}$, $\phi_0$ and $G_0$ which we have shown to be invariant under Poisson-Lie T-duality. Even better, these quantities are constant and render the field equations algebraic. This is a huge simplification when doing explicit calculation involving Poisson-Lie symmetry. For example it becomes much easier to check if an integrable deformation solved the field equations of generalized SUGRA.

There is an intricate web of relations between Poisson-Lie Symmetry/T-duality, generalized SUGRA, DFT, gauged SUGRAs and integrability. We were just able to explore a small part of it in this paper. However, there has been a lot of activity recently to go further \cite{Lust:2018jsx,Marotta:2018swj,Severa:2018pag,Crow-Watamura:2018liw,Marotta:2019wfq,Sakatani:2019jgu,Catal-Ozer:2019hxw}. Despite all these important contributions it is fair to say that there are various aspect which should be addressed in more detail. We have sorted them with increasing difficulty (at least from our perspective):
\begin{itemize}
  \item The $\mathcal{E}$-model we discussed has a non-degenerate $\mathcal{E}$. Relaxing this constraint results in the so called dressing coset construction \cite{Klimcik:1996np,Klimcik:2019kkf}. It is a generalization of standard Poisson-Lie T-duality and gives rise to double cosets as target spaces $M=\tilde H \backslash \mathdsl{D} / F$. $F$ is a second isotropic subgroup of $\mathdsl{D}$. For a $D$-dimensional physical target space, the extended space is not just doubled but its dimension is $\dim \mathdsl{D}$>2$D$. Therefore it  permits one to construct a larger number of T-dual backgrounds. An important special case is abelian/non-abelian T-duality for isometries which are acting with fixed points. It would be interesting to find a target space description which make this generalization of Poisson-Lie T-duality manifest.
   \item We have talked a lot about SUGRA where supersymmetry of course plays an important role. But we neither mentioned fermions nor supersymmetry transformations. To explore them in \DFTwzw{} would be interesting because it is known that certain abelian and non-abelian T-duality transformations break target space supersymmetry in the low energy limit\cite{Bakas:1994ba,Bergshoeff:1994cb,Sfetsos:2010uq,Kelekci:2014ima}. Intriguingly the results presented in this paper suggest that the dual target space still contains information about this seemingly ``lost'' supersymmetry. Assume that the Poisson-Lie T-duality which breaks supersymmetry only affects some of the ten space time directions in type IIA/B SUGRA. Then we are able to perform a generalized Scherk-Schwarz reduction in these directions to obtain a gauged SUGRA. More specifically, we start with a ten dimensional vacuum \footnote{This always has to be the case if at least one of the dual target spaces preserves some supersymmetry. In this case it automatically solves the field equations and gives rise to a vacuum. All dual target spaces share this property.} and reduce it to a supersymmetric vacuum of a gauged SUGRA. The generalized Scherk-Schwarz reduction just depends on the Lie group $\mathdsl{D}$ and therefore the resulting gauged SUGRA and its vacuum are identical for each of the dual target spaces. At this level we find the same amount of supersymmetry for each T-dual background. Why do we loose this information in 10D? Is it really completely lost or just hidden?
   \item As mentioned in the introduction all known (and one might conjecture that all) bosonic integrable sigma models on group manifolds can be cast as $\mathcal{E}$-models with appropriate choices for the data that defines them, for example $(\mathdsl{D}, \mathcal{E}, \eta)$. An open question is to see if this perspective can be used to further refine the landscape of integrable models and answer the additional questions: What general constraints can be placed on $\mathcal{E}$-model data that result in integrability?  Can such an analysis also be made for the case of cosets and symmetric spaces? 
  \item Exceptional field theory EFT \cite{Berman:2010is,Hohm:2013vpa} applies many of the ideas behind DFT to make U-duality manifest at the level of the low energy effective target space theory. It is formulated on an extended space which is governed by the U-duality group E$_{D(D)}$ in $D\le9$ dimensions. In this context it is natural to ask if there is a U-duality version of Poisson-Lie T-duality? There are two different ways to approach this question. One could try to formulate a $\mathcal{E}$-model for a membrane instead of a string because for M-theory from which U-duality originates, M2-branes are the fundamental objects. Exploring this approach in its outlook, \cite{Sakatani:2019jgu} concludes that it is challenging. Alternatively, one could try to formulate EFT on group manifolds to approach this problem from the target space. A small step in this direction presents \cite{duBosque:2017dfc} for $D$=$4$.
  \item All derivations in this paper were just performed in a local patch of the target space. It would be nice to better understand the global picture, too. Already at the level of non-abelian T-duality this question is subtle. Take for example the non-abelian T-dual of the compact, round three sphere. It is a non-compact target space. On the other hand there are arguments based on the AdS/CFT correspondence that suggest that this naive global picture has to be refined \cite{Lozano:2016kum}. Control over global properties is important to construct more exotic target spaces like T-folds \cite{Hull:2006va}, too. They arise if two local patches are connected by a T-duality transformation. T-folds for abelian T-duality are well studied, but there are no examples for Poisson-Lie T-duality. A first step in this interesting direction is discussed in \cite{Bugden:2019vlj}.
  \item T-duality is very closely related to the extended nature of the string and the worldsheet coupling $\alpha'$ captures its length. In the SUGRA regime we study the limit $\alpha'\rightarrow 0$ where the string is in first approximation a point particle. The actions we have studied are the leading order in a 1/$\alpha'$ expansion. Calculating subleading contributions is cumbersome. But having manifest Poisson-Lie symmetry one might hope that these calculations perhaps simplify. How could this work? One idea is to try something similar to our R/R sector approach. There, we noticed that one could do computations directly on the worldsheet. But they are very complicated and we rather exploited the symmetries we already encountered in the NS/NS sector. A Poisson-Lie group is just the classical limit of a quantum group where the deformation parameter $q\rightarrow 1$. Thus, one could try to extend the presented results to the quantum deformed version $\mathdsl{D}_q$ of $\mathdsl{D}$. At first this idea seems ad hoc, but there is a hint that suggests to incorporate quantum groups in one way or another. For a WZW-model on the Lie group $G$ at the level $k$, chiral primary fields admit the action of the quantum group $G_q$ with $q=\exp\left( i \pi / ( k + h^\vee ) \right)$ where $h^\vee$ is the dual Coexter number of $G$. These primaries are essential to derive \DFTwzw{} from CSFT in the first place. In the target space, $k$ measures the ratio between string length and the curvature of the geometry. Hence, if we send $\alpha'\rightarrow 0$ while keeping the curvature of the background finite we find the limit $q\rightarrow 1$, we are currently working in. 
\end{itemize}
This list is far from complete. But it gives an impression of the diverse research opportunities we hope to hear more about in the future.

\section*{Acknowledgements}
We thank Ctirad Klim\v{c}\'ik, Yolanda Lozano and Yuho Sakatani for comments on the draft. FH would like to thank the organisers of the Corfu Summer Institute 2018 workshop on ``Dualities and Generalised Geometries'' for inviting him to speak and the COST action MP1405 QSPACE for supporting his visit. FH also thanks the organizers and participants of the workshop ``String: T-duality, Integrability and Geometry'' in Sendai for inspiring discussions.  FH is partially supported by the Spanish Government Research Grant FPA2015-63667-.  The work of DCT is supported by a Royal Society University Research Fellowship URF 150185, and in part by the STFC grant ST/P00055X/1.  SD is supported by the FWO-Vlaanderen through aspirant fellowship and the work of DT and SD is supported in part through the projects FWO G020714N and FWO G006119N.

\bibliography{literatur}
\bibliographystyle{JHEP}

\end{document}